\documentclass[aps, reprint, superscriptaddress, showpacs, showkeys, nofootinbib]{revtex4-1} 
\usepackage{amssymb,amsfonts,amsmath}
\usepackage{graphicx}
\usepackage{epstopdf}
\usepackage{xcolor}
\usepackage[titletoc,title]{appendix}

\newcommand{\supp}{Appendix} 
\newcommand{\suppsetup}{\supp ~sec.~SI-0} 
\newcommand{\suppgauge}{\supp ~sec.~SI-1} 
\newcommand{\suppZ}{\supp ~sec.~SI-2} 
\newcommand{\suppLoop}{\supp ~sec.~SI-3} 
\newcommand{\suppGen}{\supp ~sec.~SI-4}  
\newcommand{\suppOneLoop}{\supp ~sec.~SI-6} 
\newcommand{\suppnfour}{\supp ~sec.~SI-6} 
\newcommand{\suppComplete}{\supp ~sec.~SI-5} 
\newcommand{\suppdeg}{\supp ~sec.~SI-7} 

\newcommand{\spin}{\boldsymbol{s}}

\newcommand{\vecp}{\boldsymbol{p}}
\newcommand{\spinsig}{\boldsymbol{\sigma}}
\newcommand{\hspin}{\boldsymbol{\hat{s}}}
\newcommand{\dd}{{\rm d}}
\newcommand{\hvecg}{\boldsymbol{\hat{g}}}
\newcommand{\vecg}{\boldsymbol{g}}

\newcommand{\vecempavg}{\boldsymbol{\varphi}} 
\newcommand{\empavg}{\varphi}

\begin{document}

\title{The Stochastic complexity of spin models:\\ 
Are pairwise models really simple?}

\author{Alberto Beretta}
\affiliation{The Abdus Salam International Centre for Theoretical Physics (ICTP), Strada Costiera 11, I-34014 Trieste, Italy}
\author{Claudia Battistin}
\affiliation{Kavli Institute for Systems Neuroscience and Centre for Neural
 Computation, NTNU, Olav Kyrres gate 9, 7030 Trondheim, Norway}
\author{Cl\'elia de Mulatier}
\affiliation{The Abdus Salam International Centre for Theoretical Physics (ICTP), Strada Costiera 11, I-34014 Trieste, Italy}
\author{Iacopo Mastromatteo}
\affiliation{Capital Fund Management, 23 rue de l'Universit\'e, 75007 Paris, France}
\author{Matteo Marsili}
\affiliation{The Abdus Salam International Centre for Theoretical Physics (ICTP), Strada Costiera 11, I-34014 Trieste, Italy}

\begin{abstract}
Models can be simple for different reasons: because they yield a simple and computationally efficient interpretation of a generic dataset (e.g. in terms of pairwise dependences) -- as in statistical learning -- or because they capture the essential ingredients of a specific phenomenon -- as e.g. in physics -- leading to non-trivial falsifiable predictions.
In information theory and Bayesian inference, the simplicity of a model is precisely quantified in the stochastic complexity, which measures the number of bits needed to encode its parameters. 
In order to understand how simple models look like, 
we study the stochastic complexity of spin models with interactions of arbitrary order. 
We highlight the existence of invariances with respect to bijections within the space of operators, 
which allow us to partition the space of all models into equivalence classes, in which models share 
the same complexity. We thus found that the complexity (or simplicity) of a model is not determined 
by the order of the interactions, but rather by their mutual arrangements. Models where statistical 
dependencies are localized on non-overlapping groups of few variables (and that afford predictions 
on independencies that are easy to falsify) are simple. On the contrary, fully connected pairwise 
models, which are often used in statistical learning, appear to be highly complex, because of their 
extended set of interactions. 
\end{abstract}


\maketitle

Science, as the endeavour of reducing complex phenomena to simple 
principles and models, has been instrumental to solve practical problems.
Yet, problems such as image or speech recognition and language translation 
have shown that Big Data can solve problems without necessarily 
understanding~\cite{BigDataBook,Anderson,Cristianini}.
A statistical model trained on a sufficiently large number of instances 
can learn how to mimic the performance of the human brain on these 
tasks~\cite{lecun2010convolutional,hannun5567deepspeech}.
These models are {\em simple} in the sense that they are easy to evaluate, 
train and/or to infer. They offer {\em simple} interpretations in terms of 
low order (typically pairwise) dependencies, which in turn afford an explicit 
graph theoretical representation~\cite{bishop2006pattern}.
Their aim is not to uncover fundamental laws but to ``generalize well'', i.e. 
to describe well yet unseen data. For this reason, machine learning 
relies on ``universal'' models that are apt to describe any possible data on 
which they can be trained~\cite{Wu2008}, using suitable ``regularization''
schemes in order to tame parameter fluctuations (overfitting) and achieve 
small generalization error~\cite{Goodfellow-et-al-2016}.

Scientific models, instead, are the simplest possible descriptions of 
experimental results. A physical model is a representation of 
a real system and its structure reflects the laws and symmetries of 
Nature. It predicts well not because it generalizes well, 
but rather because it captures essential features of the specific 
phenomena that it describes. It should depend on few parameters and is 
designed to provide predictions that are easy to be 
falsified~\cite{popper2005logic}. 
For example, Newton's laws of motion are consistent with momentum 
conservation, a fact that can be checked in scattering experiments.

The intuitive notion of a ``simple model'' hints at a succinct description, 
one that requires few bits~\cite{Vitanyi}. 
The {\em stochastic complexity}~\cite{rissanen1997stochastic}, derived within Minimum Description Length 
(MDL)~\cite{MDL,MDLBook}, provides a quantitative measure for 
``counting'' the complexity of models in bits. 
The question this paper addresses is: what are the features of simple models according to MDL 
and are they simple in the sense surmised in statistical learning or in physics? 
In particular, are models with up to pairwise interactions, which are frequently used
in statistical learning, simple?

We address this issue in the context of spin models, describing the statistical dependence among $n$ binary variables. 
There has been a surge of recent interest in the inference of spin 
models~\cite{Review} from high dimensional data, most of which was limited 
to pairwise models. This is partly because pairwise models allow for an intuitive graph representation of statistical dependencies.
Most importantly, since the number of $k$-variable interactions grows as $n^k$, the number of samples is hardly 
sufficient to go beyond $k=2$. For this reason, efforts to go beyond pairwise interactions have mostly focused on low order interactions (e.g. $k=3$, see \cite{Margolin2010} and references therein). 
Ref.~\cite{Ilya} recently suggested that even 
for data generated by models with higher order interactions, pairwise models 
may provide a {\em sufficiently} accurate description of the data. Within
the class of pairwise models, L1 regularization~\cite{Ranganathan} has proven
to be a remarkably efficient heuristic of model selection (but see 
also~\cite{Bulso}). 

Here we focus on the exponential family of spin models with interactions of arbitrary order. This class of models assume a sharp separation between relevant observables and irrelevant ones, whose expected value is predicted by the model. In this setting, the stochastic complexity~\cite{rissanen1997stochastic} computed within MDL coincides 
with the penalty that, in Bayesian model selection, accounts for model's complexity, under non-informative (Jeffrey's) priors~\cite{balasubramanian1997statistical}.


\subsection{The exponential family of spin models (with interactions of arbitrary order)}
Consider  $n$ spin variables $\spin=(s_1,\ldots,s_n)$, taking values $s_i=\pm 1$.
The probability distribution of $\spin$ under a model $\mathcal{M}$ belonging to the exponential family is given by:
\begin{eqnarray}
\label{eq:graph_model}
&P(\spin\,|\,\vecg,\mathcal{M})
    =\frac{1}{Z_\mathcal{M}(\vecg)}e^{\;\sum_{\mu\in \mathcal{M}} 
    g^\mu\phi^\mu(\spin)}\,,\\[4pt]
\qquad {\rm with }\;\;\;& Z_\mathcal{M}(\vecg)=\sum_{\spin} 
e^{\sum_{\mu\in \mathcal{M}} g^\mu\phi^\mu(\spin)}\,,
\end{eqnarray}
where the model $\mathcal{M}$ is identified by the set $\{\phi^\mu(\spin),~\mu\in \mathcal{M}\}$  of product 
spin operators, $\phi^\mu(\spin)=\prod_{i\in\mu} s_i\,$. 
%
Each operator $\phi^\mu(\spin)$ models the interaction that involves all the spins of 
the subset $\mu$ of the $n$ spins. 
We thus consider interactions of any arbitrary order (see~\suppsetup).
%
For instance, for pairwise interaction models, the operators~$\phi^\mu(\spin)$ are single spins $s_i$ or product of 
two spins $s_i s_j$, for $i,j \in\{1,...,n\}$. The $g^\mu$ are the conjugate
parameters\footnote{There is a broader class of models, where subsets 
$\mathcal{V}\subseteq\mathcal{M}$ of operators have the same parameter, i.e. 
$g^\mu=g^{\mathcal{V}}$ for all $\mu\in \mathcal{V}$.
These {\it degenerate models} are rarely considered in the inference 
literature. Here we confine our discussion to non-degenerate models and refer the reader to \suppdeg~for more discussion.}
that modulates the strength of the interaction associated with $\phi^{\mu}$.
Finally, the {\em partition function} $Z_\mathcal{M}(\vecg)$ ensures normalisation.
%

We remark that the models of \eqref{eq:graph_model}
can be derived as the maximum entropy distributions that are consistent with 
the requirement that the model reproduces the empirical averages of the 
operators $\phi^\mu(\spin)$ for all $\mu\in\mathcal{M}$ on a given 
dataset~\cite{jaynes1957information,Tishby}. 
In other words, empirical averages of $\phi^\mu(\spin)$ are sufficient 
statistics, i.e.\ their values are enough to compute the maximum likelihood 
parameters $\hvecg$. Therefore the choice of the operators $\phi^\mu$ in 
$\mathcal{M}$ inherently entails a sharp separation between relevant variables 
(the sufficient statistics) and irrelevant ones, which may have important consequences 
in the inference process. For example, if statistical inference assumes pairwise 
interactions, it might be blind to relevant patterns in the data resulting 
from higher order interactions. Without prior knowledge, all models $\mathcal{M}$ should
be compared.
According to MDL and Bayesian model selection (see ~\suppsetup), models should 
be compared on the basis of their maximum (log)likelihood corrected by their complexity. 
In other words, simple models should be preferred {\em a priori}.

\subsection*{Stochastic complexity} 
The complexity of a model can be defined unambiguously within 
MDL as the number of bits needed to specify {\em a priori} the parameters 
$\hvecg$ that best describe a dataset 
$\hspin = (\spin^{(1)},\dots, \spin^{(N)})$ consisting of $N$ samples 
independently drawn from the distribution $P(\spin\,|\,\vecg,\mathcal{M})$ 
for some unknown $\vecg$ (see~\suppsetup). Asymptotically for $N\to\infty$, for systems of discrete variables,
the MDL complexity is given by~\cite{Rissanen1,Rissanen2}:
\begin{equation}
	\label{MDLc}
\log \sum_{\hspin}P(\hspin\,|\,\hvecg,\mathcal{M})\,\simeq\, 
	\frac{|\mathcal{M}|}{2}\log \left(\frac{N}{2\pi}\right)+c_{\mathcal{M}}.
\end{equation}
The two terms in the r.h.s. are the {\em stochastic complexity}~\cite{rissanen1997stochastic, Myung}.
The first term, which is the basis of the Bayesian Information Criterion 
(BIC)~\cite{BIC, Myung}, captures the increase of the complexity with the 
number $|\mathcal{M}|$ of model's parameters and with the number $N$ of data 
points. This accounts for the fact that the  uncertainty in each parameter 
$\hvecg$ decreases with $N$ as $N^{-1/2}$, so its description requires 
$\sim\frac{1}{2}\log N$ bits. The second term $c_{\mathcal{M}}$ 
quantifies the statistical dependencies between the parameters, and it is given by
%
\begin{equation}
\label{eq:def:complexity}
	c_{\mathcal{M}}=\log\int d\vecg \,\sqrt{\det \mathbb{J}(\vecg)}\,,
\end{equation}
where $\mathbb{J}(\vecg)$ is the Fisher Information matrix with entries
\begin{equation}
\label{FIMm}
	J_{\mu\nu}(\vecg)=\frac{\partial^2}
		{\partial g^\mu\partial g^\nu}\log Z_{\mathcal{M}}(\vecg).
\end{equation}
The term $c_{\mathcal{M}}$ encodes for the intrinsic notion of simplicity we are 
interested in.
To distiguish these two terms, we will refer to the first as BIC term and to the second as stochastic complexity.
For an exponential family, the MDL criteria~\eqref{MDLc} coincides with the Bayesian model selection approach,
assuming Jeffreys' prior over the parameters $\vecg$~\cite{Myung, jeffreys1946invariant, Amari2} (see~\suppsetup).
Within a fully Bayesian approach, the model that maximises its posterior given the data $\hspin$, 
$P(\mathcal{M}|\hspin)$, is the one to be selected. 
Therefore, if two models have the same number of parameters (same BIC term), the simplest one, i.e. the one with the lowest stochastic complexity $c_{\mathcal{M}}$, 
has to be chosen {\em a priori}.
However, the number of possible interactions $\phi^{\mu}$ among $n$ spins is $2^n-1$, and therefore  
the number of spin models is $2^{2^n-1}$. The super-exponential growth of the number of models with the number of spins $n$ 
makes selecting the simplest model  unfeasible even for moderate $n$. Our aim is then to understand how the stochastic complexity 
depends on the structure of the model $\mathcal{M}$ and eventually provide guidelines for the search of  simpler models in such a 
huge space. 


\begin{figure}[ht]
\centering
\includegraphics[width=1\columnwidth]{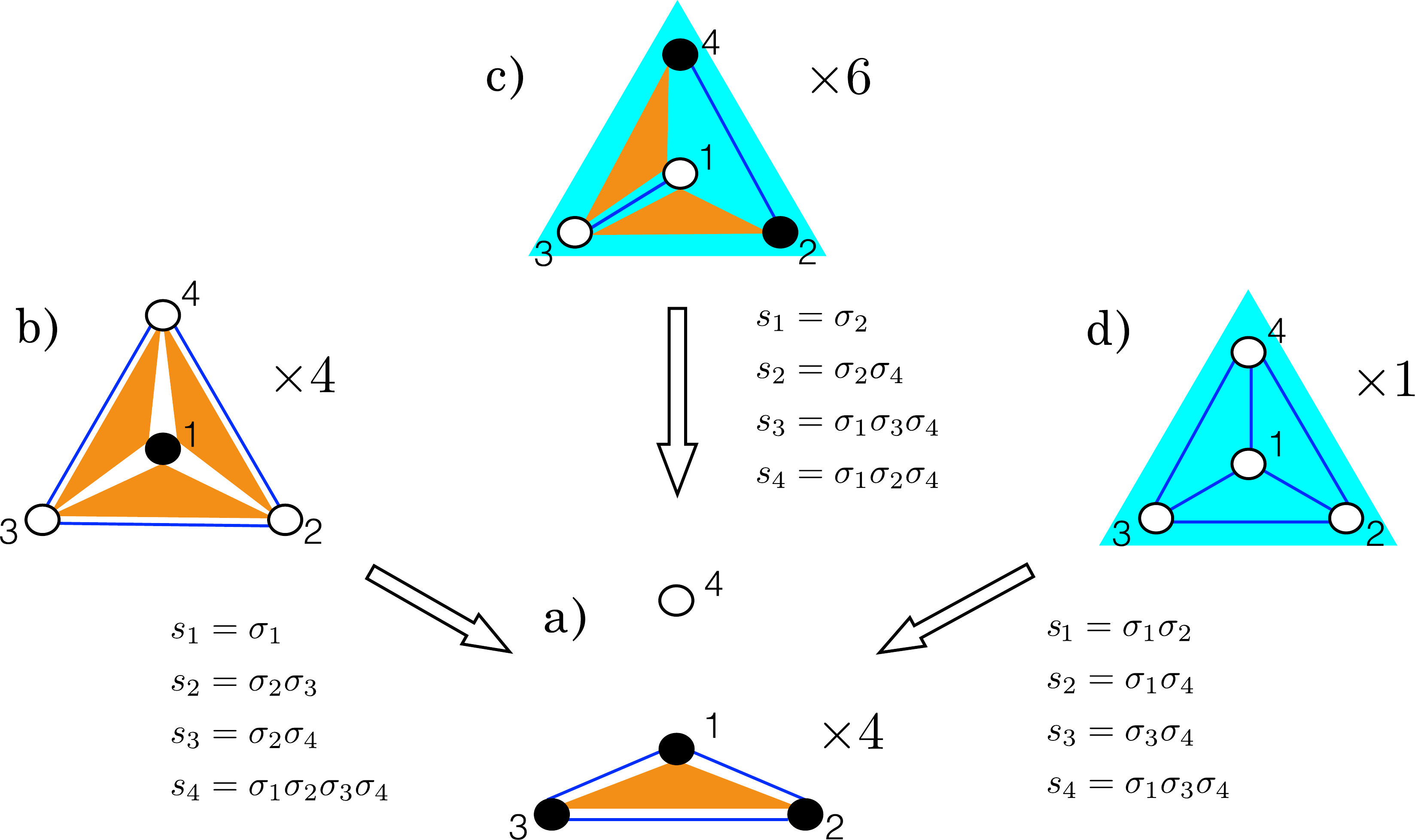} 
\caption{\label{fig1}
Example of gauge transformations between models with $n=4$ spins. Models are 
represented by diagrams (color online): spins are full dots $\bullet$ in presence of a local 
field, empty dots $\circ$ otherwise; blue lines are pairwise interactions 
($\phi^\mu=s_is_j$); orange triangles denote $3$-spin interactions 
($\phi^\mu=s_is_js_k$); and the $4$-spin interaction ($s_1s_2s_3s_4$) is a 
filled blue triangle. Note that model~a) has all its interactions grouped on 
$3$ spins; the gauge transformations leading to this model are shown along the 
arrows. All the models belong to the same complexity class, with 
$|\mathcal{M}|=7$, $\lambda=4$ and a number of independent operators 
$n_{\mathcal{M}}=3$ (e.g. $s_1, s_2$ and $s_3$ in model a) -- 
see tables in~\suppnfour). The class contains in total $15$~models, which are 
grouped, with respect to the permutation of the spins, behind the $4$ 
representatives shown here with their multiplicity~($\times m$).
}
\end{figure} 

\section*{Equivalence classes of models} 

\subsection{Gauge transformations}
Let's start by showing that low order interactions do not have a privileged 
status and are not necessarily related to low complexity $c_{\mathcal{M}}$, 
with the following argument: Alice is interested in finding which model 
$\mathcal{M}$ best describes a dataset $\hspin$; Bob is interested in the 
same problem, but his dataset $\mathbf{\hat{\spinsig}}$ is related to 
Alice's dataset by a {\em gauge transformation}. The latter is defined as 
a bijective transformation between the $n$ spin variables $\spin$ of Alice 
and those of Bob, $\spinsig=(\sigma_1, \cdots, \sigma_n)\in\{\pm 1\}^n$, that corresponds 
to a bijection from the set of all operators to itself,
$\phi^\mu(\spin)\to\phi^{\mu'}(\spinsig)$ (see the examples in 
Fig.~\ref{fig1} and \suppgauge).
This induces a bijective transformation between Alice's models and those 
of Bob, as shown in Fig.~\ref{fig1}, that preserves the number of interactions 
$|\mathcal{M}|$. Whatever conclusion Bob draws on the relative likelihood of 
models can be translated into Alice's world, where it has to coincide with 
Alice's result. It follows that two models $\mathcal{M}$ and $\mathcal{M}'$ 
related by a gauge transformation must also have the same complexity 
$c_{\mathcal{M}}=c_{\mathcal{M}'}$. In particular, pairwise interactions 
can be mapped to interactions of any order (see Fig.~\ref{fig1}), and, 
consequently, low order interactions are not necessarily simpler than higher 
order ones. 

Observe that models connected by gauge transformations have remarkably 
different structures. In Fig.~\ref{fig1}, model a) has all the possible 
interactions concentrated on 3~spins, having the properties of a 
{\it simplicial complex}\footnote{\label{simplicial_complex} 
A {\em simplicial complex}~\cite{simplicial}, in our notation, 
is a model such that, for any interaction $\mu\in \mathcal{M}$, 
any interaction that involves any subset $\nu\subseteq\mu$ of spins 
is also contained in the model (i.e. $\nu\in\mathcal{M}$).}~\cite{simplicial};
however, its gauge-transformed counterparties are not simplicial complexes.  
Model d) is invariant under any permutations of the four spins, whereas the 
other models have a lower degree of symmetry 
under permutations (see the different multiplicities in Fig.~\ref{fig1}). 


Gauge transformations are discussed in more details in~\suppgauge.
%
One can also see them as a change of the basis $\spin \to \spinsig$ in which the operators are expressed.
Counting the number of possible bases then gives us the 
number of gauge transformations (see~\suppgauge):
\begin{equation}
\label{eq:NGTmain}
	\mathcal{N}_{GT}(n)=2^{n^2}\prod_{k=1}^n\left(1-2^{-k}\right).
\end{equation}
%
%
Notice that the number of gauge transformations, \eqref{eq:NGTmain}, is much smaller than the number $2^n!$ of possible bijections of 
the set of $2^n$ states into itself. Indeed a generic bijection between the state spaces of $\spin$ and $\spinsig$ maps each product operator to one of the binary functions $f:\spinsig\to\{+1,-1\}$, which does not necessarily correspond to a product operator $\phi^{\mu}(\spinsig)$.  

\subsection{Complexity classes}
Gauge transformations allow us to divide the set of all models into 
equivalence classes, which we call {\it complexity classes}.
Models belonging to the same class are related to each other by a 
gauge transformation (that is the equivalence relation),  
and thus have the same complexity $c_{\mathcal{M}}$.
This classification suggests the presence of ``quantum numbers'' (invariants), 
in terms of which models can be classified. These invariants emerge explicitly 
when writing the cluster expansion of the partition function~\cite{
landau2013statistical,kramers1941statistics,pelizzola2005cluster} (see \suppZ):
\begin{equation}   \label{partf}
    Z_{\mathcal{M}}(\vecg)
        =
        2^n \bigg(\prod_{\mu\in \mathcal{M}}\cosh(g^{\mu})\bigg)\;
        \sum_{\mathcal{\ell}\in \mathcal{L}}\;
        \prod_{\mu\in\mathcal{\ell}}\tanh(g^{\mu})\,.
\end{equation}
The sum runs on the set $\mathcal{L}$ of all possible {\em loops} $\ell$ that 
can be formed with the operators $\mu\in\mathcal{M}$. 
A {\it loop} is any subset $\ell\subseteq\mathcal{M}$
such that $\prod_{\mu\in\ell} \phi^\mu(\spin)=1$ 
for any value of $\spin$, i.e. 
such that each spin $s_i$ occurs zero or an even number of times in this
product.
The set $\mathcal{L}$ includes the {\em empty loop} $\ell=\emptyset$. 
%
The structure of $Z_{\mathcal{M}}(\vecg)$ in \eqref{partf} depends 
on few characteristics of the model $\mathcal{M}$: the number $|\mathcal{M}|$ of operators 
(or, equivalently, of parameters) and the structure of its set of loops $\mathcal{L}$ 
(which operator is involved in which loop).
%
The invariance under gauge transformation of the complexity in~\eqref{eq:def:complexity} 
reveals itself in the fact that the partition function of models related by a gauge transformation 
have the same functional dependence on their parameters up to relabeling.

\begin{figure}[ht]
\centering
\includegraphics[width=1\columnwidth]{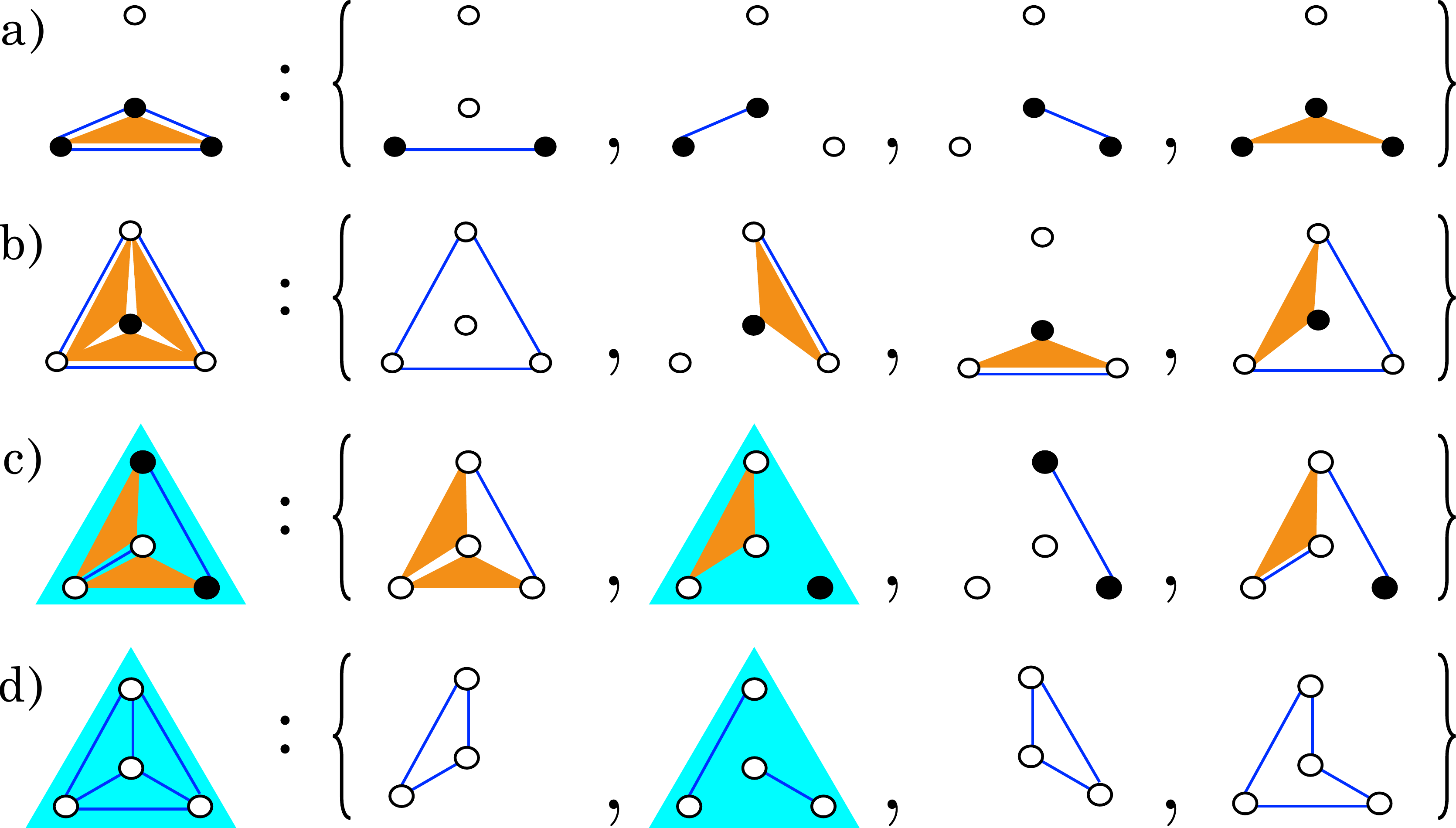}
\caption{\label{fig2}
Example of a minimal generating set of loops for each model of Fig.~\ref{fig1}.
As these models belong to the same class, their (respective) sets of loops 
have 
the same cardinality $2^\lambda$, where $\lambda=4$ is the number of generators (as shown here).
For model a), one can easily check that the $4$~loops of the set are independent,
as each of them contains at least one operator that doesn't appear
in the other $3$~loops (see \suppLoop).
%
%
%
Within each column, on the r.h.s., loops are related by the same gauge 
transformation morphing models into one another on the l.h.s. 
(i.e. the transformations displayed in Fig.~\ref{fig1}). 
This shows that the loops of these $4$ generating sets have the same structure,
which implies that the loop structure of the $4$ models is the same.
%
Any loop of a model can finally 
be obtained by combining a subset of its generating loops.
Note that the choice of the generating set is not unique.
}
\end{figure}

Let us focus on the loop structure of models belonging to the same class.
The set $\mathcal{L}$ of loops of any model $\mathcal{M}$ has the 
structure of a finite Abelian group: if $\ell_1,\ell_2\in\mathcal{L}$, 
then $\ell_1\oplus\ell_2$ is also a loop of $\mathcal{M}$, where 
$\oplus$ is the symmetric difference\footnote{
The {\em symmetric difference} of two sets $\ell_1$ and $\ell_2$ is 
the set that contains the elements that occur in $\ell_1$ but not in $\ell_2$ 
and viceversa: $\ell_1\oplus\ell_2 = (\ell_1\cup\ell_2) \setminus (\ell_1\cap\ell_2)$.
It corresponds to the XOR operator between the spins of the two loops.
} of two sets (see~\suppLoop).
As a consequence, for each model $\mathcal{M}$ one can identify a {\em minimal generating set} of $\lambda$ loops, 
such that any loop in $\mathcal{L}$ 
can be uniquely expressed as a product of loops in the minimal generating set.
Note that the choice of the generating set is not unique, though all choices have the same cardinality $\lambda$; 
Fig.~\ref{fig2} gives examples of this decomposition for the models of Fig.~\ref{fig1}.
Note also that $\ell \oplus \ell = \emptyset$ for each loop $\ell\in\mathcal{L}$. As a consequence, 
the cardinality of the loop group is $|\mathcal{L}|=2^{\lambda}$ (including the empty loop~$\emptyset$).
We found that $\lambda$ is related to the number $|\mathcal{M}|$ of operators 
of the model by $\lambda = |\mathcal{M}|-n_{\mathcal{M}}$ (see \suppLoop), where $n_{\mathcal{M}}$ is 
the number of {\it independent operators} of a model $\mathcal{M}$, i.e. the maximal 
number of operators that can be taken in $\mathcal{M}$ without forming any loop.
This implies that $\lambda$ attains its minimal value, $\lambda=0$, for models 
with only independent operators ($|\mathcal{M}| = n_{\mathcal{M}}$), 
and its maximal value, $\lambda=2^n-1-n$, 
for the {\it complete model}~$\overline{\mathcal{M}}$, that contains all the 
$|\overline{\mathcal{M}}|=2^n-1$ possible operators.
The number of independent operators $n_{\mathcal{M}}$ is preserved by gauge transformation, and, as 
the total number of operators $|\mathcal{M}|$ is also an invariant of the class, so 
is the cardinality of the minimal generating set $\lambda$.
%
%
For example, all models in Fig.~\ref{fig1} have $n_{\mathcal{M}}=3$ independent operators and $\lambda=4$ (see Fig.~\ref{fig2}).
It can also be shown that gauge transformations imply a duality relation, that associates 
to each class of models with $|\mathcal{M}|$ operators 
a class of models with the $2^n-1-|\mathcal{M}|$ complementary operators 
(see~\suppLoop).
Summarizing, the quantities $|\mathcal{M}|$ and $n_{\mathcal{M}}$, 
and the structure 
of $\mathcal{L}$ (through its generators) fully characterize a complexity class. 

\vspace{3mm}

%
%

\section*{How do simple models look like?}
\subsection{Fewer independent operators, shorter loops}
\begin{figure}[ht]
\centering
\includegraphics[width=\columnwidth]{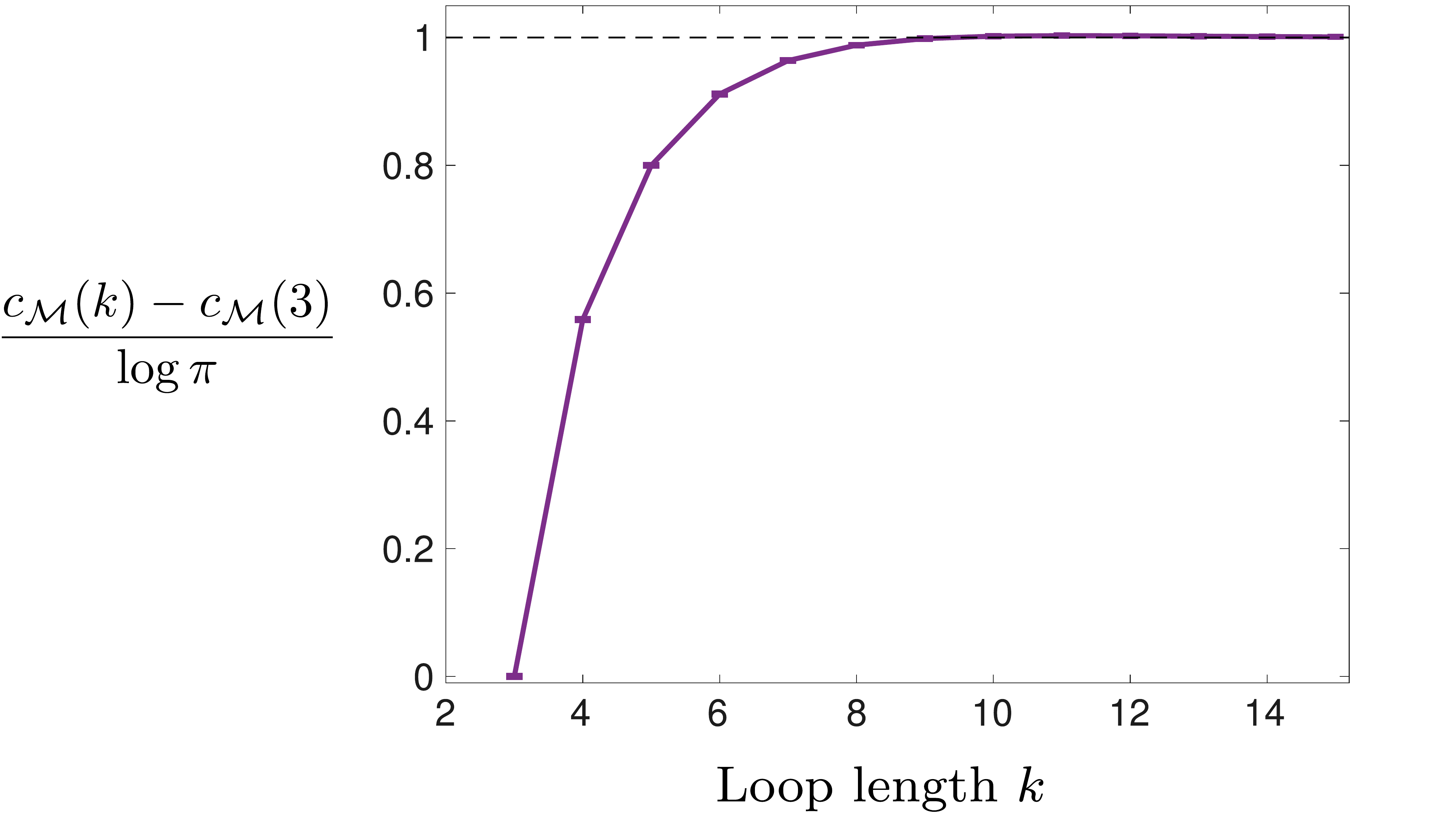}
\caption{\label{fig3} 
Complexity $c_{\mathcal{M}}(k)$ of models with a single loop of length~$k$, 
and $|\mathcal{M}|-k$ {\it free} operators, i.e. not involved in any loop. 
For $k=3$, $c_{\mathcal{M}}(3)=(|\mathcal{M}|-1)\log\pi$ can be computed 
analytically from~\eqref{eq:def:complexity}.
Values of $c_{\mathcal{M}}$ are averaged over $10^3$ numerical estimates of 
the integral in~\eqref{eq:def:complexity}, using $10^6$ Monte Carlo samples each.
Error bars correspond to their standard deviation.
}
\end{figure}

Coming to the quantitative estimate of the complexity, $c_{\mathcal{M}}$ 
generally depends on the extent to which ensemble averages 
of the operators $\phi^\mu(\spin)$ in the model $\mu\in\mathcal{M}$ constrain 
each other.
This appears explicitly by rewriting \eqref{eq:def:complexity} as an 
integral over the ensemble averages of the operators, 
$\vecempavg = \{\langle\phi^\mu\rangle, \mu\in\mathcal{M}\}$, exploiting 
the bijection between the parameters $\vecg$ and their dual 
parameters $\vecempavg$ and re-parameterization 
invariance~\cite{Amari, Amari2}:
\begin{equation}
	\label{eq:marginal_polytope}
	c_\mathcal{M}
		= \log \int_{\mathcal{F}} \dd \vecempavg 
		\sqrt{\det{\mathbb{J}(\vecempavg)}}\,,
\end{equation} 
where $\mathbb{J}(\vecempavg)$ is the Fisher Information Matrix in the 
$\vecempavg$-coordinates. The new domain~$\mathcal{F}$ of integration is 
over the values of $\vecempavg$ that can be realized in any empirical sample 
drawn from the model $\mathcal{M}$ (known in this context as 
\emph{marginal polytope}~\cite{marginal_poly}) and is related to the mutual 
constraints between the ensemble averages $\empavg^\mu$ (see~\suppGen~for more 
details).
If the model contains no loop, i.e.
$\mathcal{L} = \{ \emptyset \}$, then $J_{\mu\nu}(\vecempavg)=[1-(\varphi^\mu)^2]^{-1}\delta_{\mu\nu}$ is diagonal: 
the integral in \eqref{eq:marginal_polytope} factorizes and gives 
$c_{\mathcal{M}}={|\mathcal{M}|}\log\pi$. 
In this case, the variables $\empavg^\mu$ are not constrained at all and 
the domain of integration is $\mathcal{F}=[-1,1]^{|\mathcal{M}|}$.
If instead the model contains loops, the variables $\empavg^\mu$ become 
constrained and the marginal polytope $\mathcal{F}$ is reduced.
For example, for a model with a single loop of length three (e.g. 
$\phi^1=s_1$, $\phi^2=s_2$ and $\phi^3=s_1s_2$), 
the values of $\vecempavg$ in $[-1,1]^3$ are not all attainable, indeed
$\mathcal{F}=\{\vecempavg\in [-1,1]^3: 
~|\empavg^1+\empavg^2|-1\le\empavg^3\le 1-|\empavg^1-\empavg^2|\}$
is reduced, which decreases the complexity.
The complexity $c_{\mathcal{M}}(k)$ of models with a fixed number 
$|\mathcal{M}|$ of parameters and a single (non-empty) loop of length~$k$ is shown in 
Fig.~\ref{fig3} (see~\suppOneLoop): $c_{\mathcal{M}}(k)$ increases with $k$ and saturates at 
$|\mathcal{M}|\log\pi$, which is the value one would expect if all operators 
where unconstrained. This is consistent with the expectation 
that longer loops induce weaker constraints among the operators.
Note that the number of independent operators is kept constant here,
 equal to $n_{\mathcal{M}} = |\mathcal{M}|-1$.

\begin{figure}[h]
\centering
\includegraphics[width=\columnwidth]{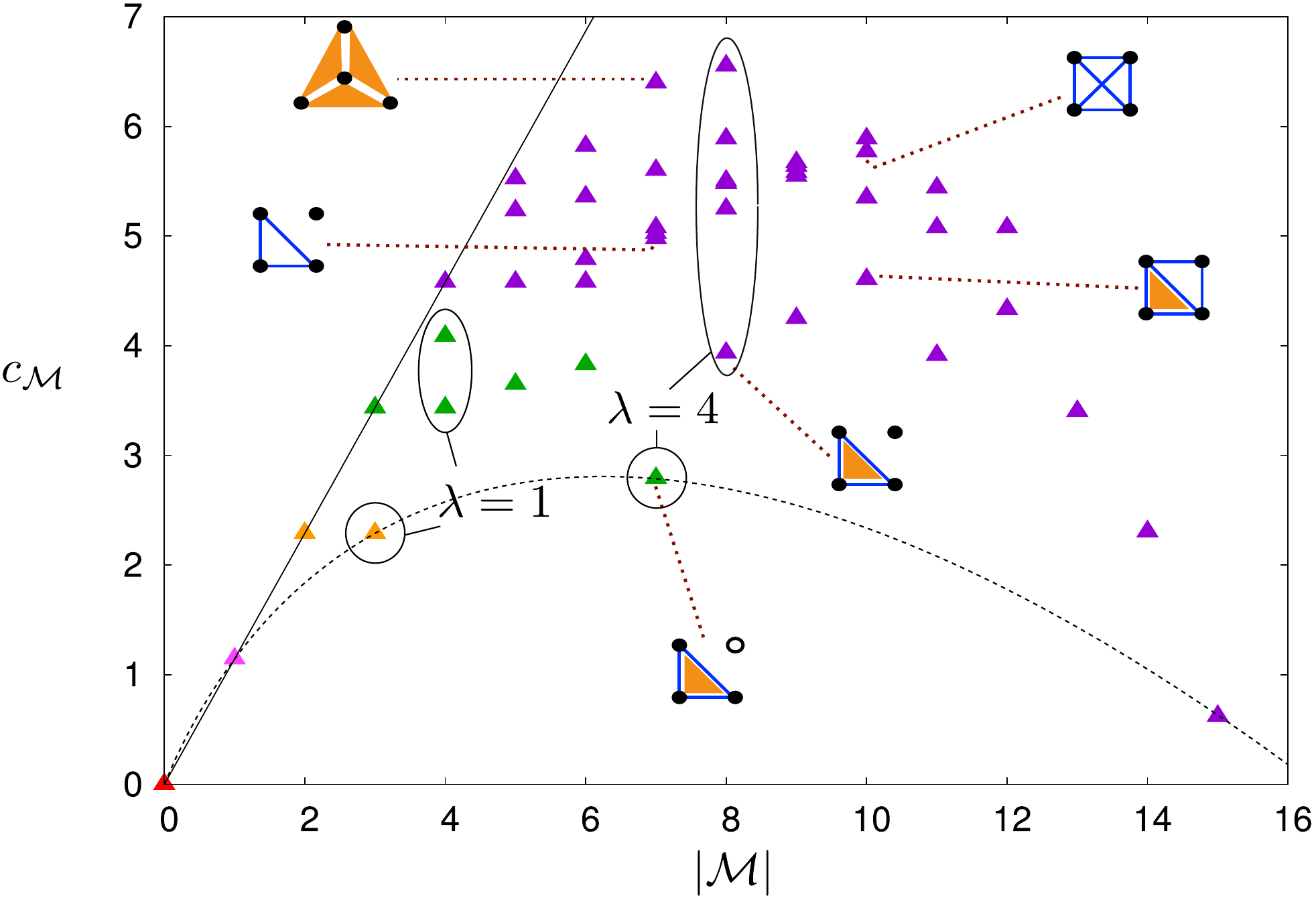}
\caption{\label{fig4} 
(color online) Complexity of models for $n=4$ as a function of the number $|\mathcal{M}|$ 
of operators: each triangle represents a class of complexity, 
which contains one or more models (see \suppOneLoop).
For each class, the value of the $c_{\mathcal{M}}$ was obtained from a 
{\it representative} of the class; some of them are shown here 
with their corresponding diagram (same notations as in Fig.~\ref{fig1}).
The triangle colors indicate the values of $n_{\mathcal{M}}$: violet for 
$n_{\mathcal{M}}=4$, green for $3$, 
yellow for $2$, pink for $1$ and red for $0$ (model with no operator).
Models on the black line have only independent operators 
($|\mathcal{M}|=n_{\mathcal{M}}$) and complexity
$c_{\mathcal{M}}=|\mathcal{M}|\,\log \pi$;  
models on the dashed curve are complete models, whose complexity is given
in~\eqref{cMcomp}.
Complexity classes with the same values of $|\mathcal{M}|$ and 
$n_{\mathcal{M}}$ have the same value of
$\lambda=|\mathcal{M}|-n_{\mathcal{M}}$, i.e. the same number of loops 
$|\mathcal{L}|$, but with a different structure.
}
\end{figure}

The single loop calculation allows computing the complexity of models with 
non-overlapping loops ($\ell\cap\ell'=\emptyset$ for all 
$\ell,\ell'\in\mathcal{L}$), for which 
$c_{\mathcal{M}}=\sum_{\ell\in\mathcal{L}}c_{\ell}$ is the sum over the 
complexity $c_{\ell}$ associated to each loop. 
In the general case of models with more complex loop structures, the explicit 
calculation of $c_{\mathcal{M}}$ is non-trivial.
Yet, the argument above suggests that, at fixed number of 
parameters~$|{\mathcal{M}}|$, $c_{\mathcal{M}}$ should increase 
with the number $n_{\mathcal{M}}$ of independent operators. 
Fig.~\ref{fig4}~summarises the results for all models with $n=4$ spins 
and supports this conclusion: for a given value of $|\mathcal{M}|$, classes 
with lower values of $n_{\mathcal{M}}$ (i.e. with less independent operators) 
are less complex. 

A surprising result of Fig.~\ref{fig4} is that $c_{\mathcal{M}}$ is not 
monotonic with the number $|\mathcal{M}|$ of operators of the model, 
increasing first with $|\mathcal{M}|$ and then decreasing. Complete models 
$\overline{\mathcal{M}}$ turn out to be the simplest (see the dashed curve 
in Fig.~\ref{fig4}). 
As a consequence, for a given $|\mathcal{M}|$, models that contain a 
complete model on a subset of spins are generally simpler than models where 
operators have support on all the spins. For instance, the complexity class 
displayed in Fig.~\ref{fig1} is the class of models with $|\mathcal{M}|=7$ 
operators that has the lowest complexity (see green triangle on the dashed 
curve in Fig.~\ref{fig4}). 

Fig.~\ref{fig4} also confirms that pairwise models are not simpler than 
models with higher order interactions. Indeed, for instance for $|\mathcal{M}|=7$, 
$c_{\mathcal{M}}$ increases drastically when changing model a) of Fig.~\ref{fig1} 
into a pairwise model 
by turning the $3$-spin interaction into an external field acting on $s_4$.
Likewise, the model with all 6 pairwise interactions for $|\mathcal{M}|=10$ 
is more complex than the one where one of them is turned into a $3$-spin 
interaction.

\subsection{Complete and sub-complete models} 
It is possible to compute explicitly the complexity of a complete model 
$\overline{\mathcal{M}}$ with $n$ spins. Indeed, there is a mapping 
$g^\mu=2^{-n}\sum_{\spin}\phi^\mu(\spin)\log p(\spin)$ between the 
$2^n-1$ parameters $g^\mu$ of $\overline{\mathcal{M}}$ and the $2^n$ 
probability $p(\spin)$, also constrained by their 
normalization~\cite{mastromatteo2013typical}.
The complexity in \eqref{eq:def:complexity} is invariant under 
reparametrization~\cite{Amari}. Re-writing this integral in terms of the 
variables $p(\spin)$ and using that 
$\det \mathbb{J}(\mathbf{p})=\prod_{\spin} 1/p(\spin)$, we find 
(see \suppComplete):
\begin{eqnarray}
\label{cMcomp}
c_{\overline{\mathcal{M}}}
    &=&\log\int_0^1\!\textrm{d}\vecp \, \delta
    \left(\sum_{\spin}p(\spin)-1\right)
    \prod_{\spin} \frac{1}{\sqrt{p(\spin)}}\,,\nonumber\\[2pt]
    &=& 2^{n-1}\log\pi-\log\Gamma(2^{n-1})\,.
\end{eqnarray}
Note that, for $n>4$, $c_{\overline{\mathcal{M}}}$ becomes negative (for 
$n=6$, $c_{\overline{\mathcal{M}}}\simeq -41.5$). 
This suggests that the class of least complex models with 
$|\mathcal{M}|$ interactions is the one that contains the model 
where the maximal number of loops are concentrated 
on the smallest number of spins.
This agrees with our previous observations on single loop models 
and sub-complete models.
On the contrary, models where interactions are distributed 
uniformly across the variables (e.g. models with only single spin operators 
for $n\ge |\mathcal{M}|$ or with non-overlapping sets of loops) have higher 
complexity. 

\subsection{Maximally overlapping loops} 
This finally leads us to conjecture that stochastic complexity is related to 
the localization properties of the set of loops~$\mathcal{L}$ (i.e. its group 
structure) rather than to the order of the interactions: models where the 
loops $\ell,\ell'\in\mathcal{L}$ have a ``large'' overlap $\ell\cap\ell'$ are 
simple, whereas models with an extended homogeneous network of interactions 
(e.g. fully connected Ising models with up-to pairwise interaction) have many 
non-overlapping loops $\ell\cap\ell'=\emptyset$ and therefore are rather 
complex.
It is interesting to note that the former (simple models) lend themselves to 
predictions on the independence of different groups of spins. 
These predictions suggest ``fundamental'' properties of the system under 
study (i.e. invariance properties, spin permutation symmetry breaking) 
and are easy to falsify (i.e. it is clear how to devise a statistical test for  
these hypotheses to any given confidence level). 
On the contrary, complex models (e.g. fully connected pairwise Ising models) 
are harder to falsify as their parameters can be adjusted to fit 
reasonably well any sample, irrespectively of the system under study. \\

\subsection{Summary}
We find that at fixed number $|\mathcal{M}|$ of operators, 
simpler models are those with fewer independent operators (i.e. smaller $n_{\mathcal{M}}$). 
For the same value of $n_{\mathcal{M}}$, models can still have different complexities.
The simpler ones are then those with a loop structure that will impose
the most constraints between the operators of the model.
More generally, we show that the complexity of a model is not defined by the order of the interactions involved, 
but is, instead, intimately connected to its internal geometry, i.e. how interactions are arranged in the model. 
The geometry of this arrangement implies mutual dependencies between interactions, that constrain the states accessible 
to the system. More complex models are those that implement fewer constraints, and can thus account for broader types 
of data. This result is consistent with the information geometric approach of Ref.~\cite{Myung}, 
which studies model complexity in terms of the geometry of the space of probability distributions\footnote{In information 
geometry~\citep{Amari2, Amari}, a model $\mathcal{M}$ defines a manifold in the space of probability distributions.
For exponential models~\eqref{eq:graph_model}, the natural metric, in the coordinates $g^\mu$, is given by the Fisher 
Information~\eqref{FIMm}, and the stochastic complexity~\eqref{eq:def:complexity} is the volume of the manifold~\cite{Myung}.}.
%
The contribution of this paper clarifies the relation between the information geometric point of view
and the specific structure of the model, i.e. the actual arrangement of its interactions.

A rough estimate of the number $N$ of data samples beyond which the complexity 
term becomes negligible in Bayesian inference can be obtained with the following argument: 
An upper bound for the complexity of models with $n$ spins and $m$ parameters is given by $m\log\pi$, 
i.e. when all operators are independent. As a lower bound, we take Eq. (\ref{cMcomp}) with $m=2^n-1$. 
This implies that an upper bound for the variation of the complexity is given by 
$\Delta c=\frac{m-1}{2}\log\pi+\log\Gamma\left(\frac{m+1}{2}\right)$. When this is much smaller than 
the BIC term, the stochastic complexity can be neglected. For large $m$ this implies $N\gg m$, which may be 
relevant for the applicability of fully connected pairwise models ($m\simeq n^2/2$) in typical cases,
for instance when samples cannot be considered as independent observations from 
a stationary distribution (see~\cite{Bulso}).

\section*{Conclusion}
As pointed out by Wigner~\cite{Wigner} long ago, the unreasonable effectiveness of mathematical models 
relies on isolating phenomena that depend on few variables, whose mutual variation is described by {\em simple models} 
and is independent of the rest. Remarkably we find that, for a fixed number of spin variables and parameters, simple models, 
according to MDL, are precisely of this form: statistical dependencies are concentrated on the smallest subset of variables 
and these are independent of all the rest. 

Such simple models are not optimal to generalize, i.e. to describe generic statistical dependencies, rather they are easy to falsify. 
They are designed for spotting independencies that may hint at deeper principles (e.g. symmetries or conservation laws) that may 
``take us beyond the data''~\footnote{In his response to Ref.~\cite{Anderson} 
on {\protect \tt edge.org}, W.D. Willis observes that ``Models are 
interesting precisely because they can take us beyond the data''.}. On the contrary, fully connected pairwise models appears to be rather complex. 
This, we conjecture, is the origin of {\em pairwise sufficiency} \cite{Ilya} that makes them so successful to describe a wide variety of data from 
neural tissues \cite{Bialek1} to voting behaviour \cite{BialekUSSC}. 

On the other hand, pairwise interactions play a special role in our understanding of phenomena 
as they allow to reduce statistical dependencies into {\em direct interactions} between variables. 
Therefore it would be important to identify methods to quantitatively assess when a dataset is genuinely 
described by pairwise interactions.
The results of this paper allow one to address this issue by comparing inference with pairwise models 
to inference with models obtained via their gauge transformations. 
Since the latter preserve the number of interactions and the stochastic complexity, transformed models 
have the same flexibility in terms of generalisation. 
For the same reason, the comparison between pairwise models and their gauge transformed ones can be done 
on the basis of likelihood alone. 

In conclusion, our results suggest that when data are scarce and high dimensional, Bayesian inference should 
privilege simple models, i.e. those with small stochastic complexity, over more complex ones, such as fully 
connected pairwise models that are often used~\cite{Review, Bialek1,BialekUSSC}. A full Bayesian model selection 
approach is hampered by the calculation of the stochastic complexity that is a daunting task. Developing approximate 
heuristics for accomplishing this task is a challenging future avenue of research.

\begin{acknowledgments}
C. B. acknowledges financial support from the Kavli Foundation and the 
Norwegian Research Council's Centre of Excellence scheme (Centre for Neural 
Computation, grant number 223262). A. B. acknowledges financial support from 
International School for Advanced Studies (SISSA).
\end{acknowledgments}

\bibliographystyle{apsrev4-1}
\nocite{wainwright2003variational}
\bibliography{pnas-sample}

\end{document}


\title{Supporting Information Appendix for:\\
"The Stochastic complexity of spin models:\\ 
Are pairwise models really simple?"} 
\author{Alberto Beretta}
\affiliation{The Abdus Salam International Centre for Theoretical Physics (ICTP), Strada Costiera 11, I-34014 Trieste, Italy}
\author{Claudia Battistin}
\affiliation{Kavli Institute for Systems Neuroscience and Centre for Neural
 Computation, NTNU, Olav Kyrres gate 9, 7030 Trondheim, Norway}
\author{Cl\'elia de Mulatier}
\affiliation{The Abdus Salam International Centre for Theoretical Physics (ICTP), Strada Costiera 11, I-34014 Trieste, Italy}
\author{Iacopo Mastromatteo}
\affiliation{Capital Fund Management, 23 rue de l'Universit\'e, 75007 Paris, France}
\author{Matteo Marsili}
\affiliation{The Abdus Salam International Centre for Theoretical Physics (ICTP), Strada Costiera 11, I-34014 Trieste, Italy}

\setcounter{section}{-1}
\renewcommand{\thesection}{SI-\arabic{section}}
\renewcommand{\thesubsection}{SI-\arabic{section}.\arabic{subsection}}
\setcounter{equation}{0}

\maketitle

\section{General framework -- Spin models}\label{SI-0}

\subsection{Spin operators} 
Let us consider the system of $n$ spin variables, $\spin=(s_1, \,\cdots, s_n)$, that take random values $s_i=\pm 1$. In order to account for 
interactions of any order, we associate to any interaction involving a subset $\mu$ of spins a {\it spin operator} $\phi^\mu(\spin)$ defined 
as the product of all the spins in $\mu$:
\begin{equation}
    \label{eq:def:operators}
    \phi^{\mu}(\spin)=\prod_{i\in \mu} s_i \,, 
\end{equation}
which takes value in $\{+1,-1\}$.
By definition, the total number of these operators corresponds to the number of possible interactions in the $n$-spin system, 
i.e. to the number of possible subsets of $\{s_1, \dots, s_n\}$, empty set excluded, which is $2^n-1$.
In the following we simplify the notation of the operator label $\mu$ by using an integer, $\mu\in\{1,\dots,2^n-1\}$, 
whose binary representation directly identifies the spins that belong to the set $\mu$~\footnote{\label{Note1}The spin {$\protect s_i$} belongs to {$\protect\mu $} if the {$\protect i$}th digit (starting from the right) in the binary representation of {$\protect\mu $} is ``{$\protect 1$}''. For example, {$\protect\phi^5(s)=s_1 s_3$}, since the binary representation of {$\protect 5$} is {$0\protect \dots  0101$}, with ``{$\protect 1$}'' only in the {$\protect 1$}rst and {$\protect 3$}rd positions along the binary string of length {$\protect n$}.}.
For instance, the operator $\phi^{1}(\spin)=s_1$ is associated with a field acting on~$s_1$, and $\phi^{7}(\spin)=s_1s_2s_3$ with a three body interaction.
These {\it spin operators} are the building blocks of the models. Note that, these operators verify:
\begin{equation}
\label{eq:sum_phi}
    \sum_{\spin\in\mathcal{S}} \phi^\mu(\spin)=0\,, \qquad \mu\in\{1,\,\dots,2^n-1\}\,,
\end{equation}
where $\mathcal{S}=\{-1, 1\}^n$, and the sum over $\spin\in\mathcal{S}$ denotes the sum over all possible configurations of the spins.

\subsection{\textit{Complete} set of spin operators}
We define the set $\Omega_n=\{\phi^\mu(\spin)\}_{\mu\in\{0, \dots, 2^n-1\}}$ of all the spin operators 
built with $n$ spins, including also the operator $\phi^0(\spin)=1$ (which is not associated with any interaction). 
By definition, the cardinality of $\Omega_n$ is $|\Omega_n|=2^n$.
The set $\Omega_n$ is (called) {\it orthogonal} and {\it complete} as its operators verify respectively the relations~\cite{mastromatteo2013typical}: 
\begin{equation}
\label{orth}
    \langle \phi^\mu(\spin), \phi^\nu(\spin)\rangle=\frac{1}{2^n}\sum_{\spin\in\mathcal{S}} \phi^\mu(\spin)\,\phi^\nu(\spin)=\delta_{\mu,\nu}
        \qquad {\rm and} \qquad
    \frac{1}{2^n}\sum_{\mu=0}^{2^n-1}\phi^\mu(\spin)\,\phi^\mu(\spin') = \delta_{\spin,\spin'}.
\end{equation}
The first relation defines an inner product $\langle \cdot, \cdot\rangle$ over the space of operators $\Omega_n$.
The relation derives from the fact that the product of two operators of $\Omega_n$ is also an operator of $\Omega_n$:
\begin{align}
\label{eq:prod_op}
    \phi^\mu(\spin)\,\phi^\nu(\spin)=\phi^{\mu\oplus\nu}(\spin)\,,
\end{align}
where, in the binary representation of $\mu$ and $\nu$, $\oplus$ is the XOR bitwise operation. 
Using the property \eqref{eq:sum_phi} and observing that $\phi^\mu(\spin)\phi^\mu(\spin)=\phi^0(\spin)=1$ gives the first relation.
The second relation is an immediate consequence of the fact that, for the set of monomials $\Omega_n$, one has
\begin{align}
\label{eq:completeness}
    \sum_{\mu=0}^{2^n-1}\phi^\mu(\spin)\,\phi^\mu(\spin') 
    &= \sum_{\alpha_1=0,1} \cdots \sum_{\alpha_n=0,1} \, \prod_{i=1}^n (s_i s'_i)^{\alpha_i}\,,
\end{align}
which is always equal to zero, unless all $s_i$ are equal to $s'_i$. In this latter case the sum yields the $2^n$ factor that allows to recover \eqref{orth}. 
The orthogonality and completeness properties~\eqref{orth} allow us to express~\footnote{
Note that the set of functions $F:\mathcal{S}\to \mathbb{R}$ is a vector space provided with the sum $(F+G)(\spin)=F(\spin)+G(\spin)$ and the multiplication by scalar in
$\mathbb{R}$. It is also provided with the scalar product $<F,G>=\protect\frac{1}{2^n}\protect\sum_{\spin}F(\spin)G(\spin)$ and $\Omega_n$ as an orthonormal basis.}
any function $F(\spin)$ as a linear combination of operators~\cite{mastromatteo2013typical} $\phi^\mu(\spin)$, i.e. 
\begin{equation}
\label{ }
    F(\spin)=\sum_{\mu\in\Omega_n} f^\mu\phi^\mu(\spin),
    \qquad {\rm and} \qquad
    f^\mu=\langle F(\spin), \phi^\mu(\spin)\rangle = \frac{1}{2^n}\sum_{\spin\in\mathcal{S}} \phi^\mu(\spin) F(\spin).
\end{equation}

\subsubsection*{\textbf{Generating set of $\Omega_n$ and independent operators}}
In the following, we will call {\it generating set} of $\Omega_n$ a set of $n$ spin operators that can fully generate $\Omega_n$, such as the set $\{s_1, \dots, s_n\}$.
We also define the notion of {\it set of independent operators}~\footnote{
Note: with the definition of {\it loops}, that will be introduced in \ref{suppZ}, a set of {\it independent operators} is a set of operators that doesn't form any loop (but the {\it empty loop}).
}
as a set $\mathcal{I}$ verifying that the product of all the operators of any subset of $\mathcal{I}$ is always different from $\phi^0(\spin)=1$.
Formally, any set of $n$ {\it independent operators} of $\Omega_n$ is a {\it generating set} of $\Omega_n$.
By definition, a generating set of $\Omega_n$ cannot include the identity operator $1$.\\

Mathematically, the set $\Omega_n$, associated with the multiplication operation $(\Omega_n, \cdot)$, forms a finite Abelian group with identity element $\phi^0(s)=1$ generated by a minimal set of $n$ generators of order 2 ($\Omega_n= {\mathbb{Z}_2}^n$).


\subsection{Spin models} \label{SI-02}
A {\it model} $\mathcal{M}$ is defined in terms of a subset $\mathcal{M}\subseteq\Omega_n\backslash \{\phi^0\}$ of operators~\footnote{We will use the same notation $\mathcal{M}$ for the model, the subset of operators, or the subset of values of $\mu$ that identify the operators in the model.}.
These define a probability distribution of the vector $\spin=(s_1,\ldots,s_n)$ of spin variables:
\begin{equation}
\label{eq:graph_modelA}
P(\spin\,|\,\vecg,\mathcal{M})=\frac{1}{Z_\mathcal{M}(\vecg)}e^{\sum_{\mu\in \mathcal{M}} g^\mu\phi^\mu(\spin)}\qquad\qquad{\rm where}\qquad Z_\mathcal{M}(\vecg)=\sum_{\spin\in\mathcal{S}} e^{\sum_{\mu\in \mathcal{M}} g^\mu\phi^\mu(\spin)}\,,
\end{equation}
where the vector $\vecg=\{g^\mu,~\mu\in \mathcal{M}\}$ are the conjugate parameters:
each parameter $g^\mu$ is a real variable that modulates the strength of the interaction associated with the operator $\phi^\mu(\spin)$. We shall refer to the model $\bar{\mathcal{M}}=\Omega_n\backslash \{\phi^0\}$ with all operators as the {\em complete model}.
%
Models can be {\it degenerate} (several operators are mapped to the same parameter) or not.

\subsubsection*{\textbf{Non-degenerate models}} {\it Non-degenerate models} are those for which each operator $\phi^\mu(\spin)$ is assigned a different parameter 
$g^\mu$.
For instance, model a) in Fig.~\FigOne~ of the main text involves $|\mathcal{M}|=7$ interactions, mapping the 7 operators $\mathcal{M}=\{s_1, s_2, s_3, s_1s_2, s_1s_3, s_2s_3, s_1s_2s_3\}$ onto the 7 parameters $\vecg=\{g^1,g^2,g^4,g^3,g^5,g^6,g^7\}$ (using the binary representation\textsuperscript{\ref{Note1}} of $\mu$).
The number of different non-degenerate models with $n$ spins grows 
superexponentially in $n$:
\begin{align}
    \mathcal{N}_n = 2^{|\Omega_n\backslash\{1\}|} = 2^{2^n -1}\,.
\end{align}
To give an idea: $\mathcal{N}_2=8$, $\mathcal{N}_3=128$, $\mathcal{N}_4=32 768$, $\mathcal{N}_5\simeq 2\cdot 10^9$.
In the main paper and in most of the supplemental material we shall focus on non-degenerate models. \\

\subsubsection*{\textbf{Degenerate models}} 
For completeness we also define {\em degenerate models}, which are discussed in section~\ref{SI-7}. In a {\it degenerate model}, each parameter can be associated to one or more interactions. For example, the mean field Ising model is a degenerate model with only 2 parameters, $h$ and $J$; the connection with the $g^\mu$ notation 
reads:
\[
g^\mu=\left\{\begin{array}{lll}
    h\, \qquad & \hbox{for~~all~~}\;\mu=2^k &{\rm with~~}k\in[0,n-1]\,, \\
    J\, & \hbox{for~~all~~}\;\mu=2^k+2^{k^\prime}\,&{\rm with~~}k>k^\prime {\rm ~~and~~} (k,k^\prime)\in[0,n-1]^2\,, \\
    0\, & \hbox{otherwise}&,
\end{array}\right.
\]
where we used the binary representation of the set $\mu$. 
To work with degenerate models, it is convenient to introduce a more general notation,
in which a model $\mathcal{M}$ is defined by a set of $|\mathcal{M}|$ operators, $\boldsymbol{\phi}=\{\phi^\mu\}_{\mu\in\mathcal{M}}$, a set of $m$ parameters $\vecg=\{g^i\}_{i\in\{0,\dots, m\}}$, and a rectangular (mapping) matrix $\matQ$ of size $|\mathcal{M}|\times m$ that maps each operator of $\boldsymbol{\phi}$ to one parameter of $\vecg$:
\begin{align}
U_{ij}=
    \begin{cases}   
        \; 1 &\qquad \textrm{if }\phi^j\textrm{ is parameterised by }g^i\,,\\
        \; 0 &\qquad {\rm otherwise}\,.
    \end{cases}
    \label{eq:matrixQ}
\end{align}
For non-degenerate models, $\matQ$ is simply the $|\mathcal{M}|\times|\mathcal{M}|$ identity matrix.
By definition, each column of $\matQ$ contains a single $1$, whereas the sum of each line $i$ gives the degeneracy $\alpha_i=\sum_{j} U_{ij}$ of the parameter $g^i$.
Note that $\sum_{ij} U_{ij}=|\mathcal{M}|$. 
The extension to such {\em degenerate} models is natural when operators $\mu\in \mathcal{V}$ are of the same order~\footnote{However, let us remark that this symmetry is not preserved under the {\it gauge transformations} that will be introduced later, because two operators of the same order can be mapped to operators of different orders (see~\ref{SI-1}).}.
The number of possible degenerate models, where interactions of the same order may be assigned the same parameter, grows much faster than $\mathcal{N}_n$ with $n$:
\[
\mathcal{N}^{deg}_n = \prod_{j=1}^n \; B_{{n \choose j} + 1}\;,
\]
where $B_{m}$ is the number of partitions of a set with $m$ elements, known as Bell number. For instance,
$\mathcal{N}^{deg}_2=10$, $\mathcal{N}^{deg}_3=450$, $\mathcal{N}^{deg}_4 = 2.371.408$ and $\mathcal{N}^{deg}_5 \simeq 38 \cdot 10^{15}$.

\subsection{The stochastic complexity and Bayesian Model Selection}
In this section we recall the relation between the {\it stochastic complexity} defined in the context of Minimum Description Length and the {\it geometric complexity} obtained from a Bayesian approach. 
We refer to Refs. \cite{rissanen1996fisher,*Myung2000counting} for a more complete treatment.\\

Bayesian model selection dictates that, given a dataset $\hspin=(\spin^{(1)},\ldots,\spin^{(N)})$ of $N$ observed configurations~$\spin^{(i)}\in\mathcal{S}$, each model should be assigned a posterior probability
\begin{equation}
    \label{BMS}
    P(\mathcal{M}\,|\,\hspin)=\frac{P(\hspin\,|\,\mathcal{M})\,P_0(\mathcal{M})}{\sum_{\mathcal{M}'}P(\hspin\,|\,\mathcal{M}')\,P_0(\mathcal{M}')},
\end{equation}
according to Bayes' rule. 
Here $P_0(\mathcal{M})$ is the prior probability on the model $\mathcal{M}$ and the sum in the denominator runs on all models $\mathcal{M}'$ that are considered. In Eq. (\ref{BMS}), $P(\hspin|\mathcal{M})$ is the so-called evidence that is computed by integrating the likelihood over the parameters. In the case where $\spin^{(i)}$ are i.i.d., drawn from a distribution $P( \spin^{(i)}\,|\,\vecg,\mathcal{M})$, this reads:
\begin{align}
    \label{evidence}
    P(\hspin\,|\,\mathcal{M})
        &= \int\dd\vecg \;\prod_{i=1}^N P( \spin^{(i)}|\,\vecg,\mathcal{M})\; P_0(\vecg|\mathcal{M})\,
\end{align}
where $P_0(\vecg|\mathcal{M})$ is the prior distribution on the parameters $\vecg$ of model $\mathcal{M}$. 
For spin models, the probability $P( \spin\,|\,\vecg,\mathcal{M})$ is given by Eq. (\ref{eq:graph_modelA})
and the evidence becomes:
\begin{align}
    \label{evidence:spinM}
    P(\hspin\,|\,\mathcal{M})
        &= \int\dd\vecg \;
        \e^{\,\textstyle N\left[\vecempavg(\hspin)\cdot\vecg - \log Z_{\mathcal{M}}(\vecg) \right]\,}\,
        P_0(\vecg\,|\,\mathcal{M})\,,
\end{align}
where $\vecempavg(\hspin)$ is a vector with $|\mathcal{M}|$ elements, containing the empirical averages of the operators $\phi^{\mu}$ over the measured data $\hspin$:
\begin{align}\label{eq:empirical_av}
    \empavg^\mu(\hspin) = \frac{1}{N} \sum_{i=1}^{N} \phi^{\mu}(\spin^{(i)})\,, \qquad {\rm for}\;\; \mu\in\mathcal{M}\,.
\end{align}
The log-likelihood, $\log P(\hspin\,|\,\vecg, \mathcal{M}) = N\,[\vecempavg(\hspin)\cdot\vecg - \log Z_{\mathcal{M}}(\vecg)]$, is a convex function of $\vecg$ and it has a unique maximum for the values of the parameters $\vecg=\hvecg$ that are the solution of the set of equations: 
\begin{align}\label{eq:maxlikelihood}
    \ensavg^\mu(\hvecg) = \empavg^{\mu}(\hspin)
        \qquad
        \textrm{for all }\,\mu\in\mathcal{M}\;,  
\end{align}
where
\begin{align}
\ensavg^\mu(\vecg) = \frac{\partial \log Z_{\mathcal{M}}(\vecg)}{\partial g^\mu}
    = \sum_{\spin\in\mathcal{S}} \phi^{\mu}(\spin)\; P(\spin\,|\,\vecg, \mathcal{M})\,,
    \label{eq:ens_averages}
\end{align}
denotes the ensemble average of the operator $\phi^\mu(\spin)$ under the model specified by $\vecg$.
In other words, at $\vecg=\hvecg$, the ensemble average of each operator $\phi^\mu$ of $\mathcal{M}$ is equal to its empirical average $\empavg^\mu$ 
over the measured data $\hspin$. For large $N$, the integral is sharply dominated by the maximum and it can be estimated by the Saddle-point method, expanding $\vecempavg(\hspin)\cdot\vecg - \log Z_{\mathcal{M}}(\vecg)$ to second order about $\hvecg$.  
This shows that, for large $N$, Eq.~(\ref{evidence:spinM}) is well approximated by:
\begin{align}\label{eq:likelihood}
    \log P(\hspin\,|\,\mathcal{M}) \cong 
        \log P(\hspin\,|\,\hvecg, \mathcal{M})\,
        -\,\frac{|\mathcal{M}|}{2} \log\left(\frac{N}{2\pi}\right)\,
        -\,c^{\rm \scriptscriptstyle BMS}_{\mathcal{M}}
        +\,O\left(\frac{1}{N}\right)\,,
\end{align}
where $c^{\rm \scriptscriptstyle BMS}_{\mathcal{M}}$ is a geometric complexity term~\cite{Myung2000counting} arising from the Gaussian integration:
\begin{align}
    c^{\rm \scriptscriptstyle BMS}_{\mathcal{M}}=\log\left[\frac{\sqrt{\det
                \mathbb{J}(\hvecg)}}{P_0(\hvecg\,|\,\mathcal{M})}
                \right]\,,
\label{eq:BMSc}
\end{align}
and $\mathbb{J}(\vecg)$ is the Hessian of the log-likelihood, which in this case coincides with the Fisher Information matrix defined in Eq.~(\EqFIMm) of the main text.\\

Minimum Description Length instead approaches the problem of model complexity from an apparently different angle. Imagine we run a series of experiments that generate a sample $\hspin$ of $N\gg 1$ observations of a system. We model the outcome of the experiment as $N$  i.i.d. drawn from a model $P(\spin\,|\,\vecg,\mathcal{M})$ for unknown parameters $\vecg$ (imagine the situation where we run the experiment precisely because we want to infer the parameters $\vecg$). How much memory storage should be set aside {\em before} running the experiment? If we knew the parameters $\hvecg$ the solution is given by (minus) the log-likelihood $-\log P(\hspin\,|\,\hvecg,\mathcal{M})$ where $P(\hspin|\ldots)=\prod_{i=1}^N P(\spin^{(i)}|\ldots)$. In the absence of this information, the problem can be cast as a minimax problem (we refer to \cite{grunwald2007minimum} for details), i.e. to find the best possible coding $\bar P(\hspin)$ in the case where Nature choses the worst possible sample $\hspin$. The solution is the {\em normalised maximum likelihood}
\begin{equation}
\label{UniversalCode}
\bar P(\hspin)=\frac{P(\hspin\,|\,\hvecg(\hspin),\mathcal{M})}{\sum_{\hspin^\prime}P(\hspin^\prime\,|\,\hvecg(\hspin^\prime),\mathcal{M})}.
\end{equation}
From this, it is clear that the additional memory space that is needed to describe the model and the parameters is given by the log of the denominator of Eq. (\ref{UniversalCode}), which is the l.h.s. in Eq.~(\EqMDLc) of the main text. 
In order to derive the r.h.s. of Eq.~(\EqMDLc) of the main text, consider the expansion: 
\begin{equation}
\label{UC1}
\int d\vecg \, P(\hspin\,|\,\vecg,\mathcal{M})\,f(\vecg) \simeq  P(\hspin\,|\,\hvecg,\mathcal{M})\,f(\hvecg)\,\frac{(2\pi/N)^{|\mathcal{M}|/2}}{\sqrt{\det \mathbb{J} (\hvecg)}}\, \left[1+O(1/N)\right]
\end{equation}
that arises from performing the integral by saddle point around the maximum likelihood parameters $\hvecg(\hspin)$ which depend on the data $\hspin$. 
In Eq.~(\ref{UC1}),
the matrix~$\mathbb{J}$ is, in general, the Hessian of the  likelihood at $\hvecg$. Yet, for exponential models, the Hessian $\mathbb{J}$ does not depend on the data, and it coincides with the Fisher Information matrix. 
Taking $f (\vecg)=\sqrt{\det \mathbb{J} (\vecg)}$, summing over all samples $\hspin$ in Eq. (\ref{UC1}) and taking the limit $N\to\infty$, one finds
\begin{equation}
\label{mdlc}
\int d\vecg \sqrt{\det \mathbb{J} (\vecg)}=\lim_{N\to\infty}\left(\frac{2\pi}{N}\right)^{|\mathcal{M}|/2}\sum_{\hspin} P(\hspin\,|\,\hvecg,\mathcal{M})=e^{c_{\mathcal{M}}}\,,
\end{equation}
which is Eq.~(\EqMDLc) of the main text, and where $c_{\mathcal{M}}$ is given by Eq.~(\Eqdefcomplexity) of the main text. \\

As observed in Ref.~\cite{Myung2000counting}, the choice of {\it Jeffreys' priors}~\cite{jeffreys1946invariant}
\begin{equation}
    \label{cM}
    P_0(\vecg\,|\,\mathcal{M})=
        \frac{\sqrt{\det \mathbb{J}(\vecg)}}{\int\dd \vecg'\,\sqrt{\det \mathbb{J}(\vecg')}}\;
\end{equation}
in Eq.~(\ref{eq:BMSc}) makes the geometric complexity $c^{\rm \scriptscriptstyle BMS}_{\mathcal{M}}$ of the Bayesian approach 
coincide with the stochastic complexity $c_{\mathcal{M}}$ (see Eq.~(\Eqdefcomplexity) of the main text) prescribed 
by Minimum Description Length~\cite{rissanen1996fisher, rissanen1997stochastic}. 
This choice for the prior seems natural (in absence of any information on the values of $\vecg$), 
as it corresponds to assuming an {\em a priori} uniform distribution in the space of samples~\cite{Myung2000counting}.
We will see that this choice of prior has also an interesting property, as it is invariant under re-parametrisation, which will 
lead to the definition of class of complexity.

\section{Gauge transformations (GT)}\label{app:nb_GT}\label{SI-1}
\subsection{Definition}
Any generating set $\sigmaset=\{\phi^{\nu_1},\ldots,\phi^{\nu_n}\}$ of $\Omega_n$ induces a bijection $\spin\to\sigmaset(\spin)$ on the set of configurations $\mathcal{S}$ and on the set $\Omega_n$ of operators. Indeed
\[
\phi^\mu(\sigmaset)=\prod_{i\in\mu}\phi^{\nu_i}(\spin)=\phi^{\mu'}(\spin),\qquad \mu'=\oplus_{i\in\mu}\nu_i
\]
where $\oplus_{i\in\mu}\nu_i$ is the bitwise XOR of the binary representation of the integers $\nu_i$ for all $i\in\mu$. We call such a bijection a {\it gauge transformation}~\footnote{Note that the set of configuration $\spin=(s_1,\ldots,s_n)$ is itself a set of operators in $\Omega_n$. So strictly speaking GTs can be defined as bijections from $\Omega_n$ to itself defined by the operation $\mu\to\mu'=\oplus_{i\in\mu}\nu_i$ for any particular choice of $\nu_i$ that realises a generating set of $n$.}. In other words, these are transformations that map the set of $n$ generators $\{s_1, \dots, s_n\}$ of $\Omega_n$ to another set of generators of $\Omega_n$, i.e. a set of $n$ {\it independent operators} of $\Omega_n$ (see definitions in~\ref{SI-0}).
A GT {\em preserves the structure} of $\Omega_n$ in the sense that any operator in the old basis is mapped into a distinct operator in the new one. A transformation that maps $(s_1, \dots, s_n)$ to a set of $n$ non-independent operators will not preserve its structure. Indeed it maps $\Omega_n$ to a strict subset of $\Omega_n$, with $n^{\prime}<n$  independent generators. Combining them can generate only $2^{n^{\prime}}$ operators, which means that some operators of $\Omega_n$ will not occur in the new basis.
Note also that the operator $\phi^0(\spin)=1$ is invariant under GTs.

Mathematically, these transformations are the {\it automorphisms} of the group $(\Omega_n, \cdot)$.

\subsection{Number of gauge transformations for a system with $n$ spins}\label{SI-2.1}
The total number of these transformations corresponds to the number of possible sets of generators of $\Omega_n$.
There are exactly
\begin{align}
    \dbinom{|\Omega_n|-1}{n}\times n! = \prod_{i=1}^n \, (2^n-i) = \left(2^n\right)^n \, \prod_{i=1}^n \left(1-\frac{i}{2^n}\right)\simeq \left(2^n\right)^n\left(1+O(n^2/2^{n})\right)
\end{align}
possible ways to sample a set of $n$ operators among $\Omega_n\backslash \{1\}$ \footnote{
We recall that a generating set of a $\Omega_n$ cannot include the identity operator (see~\ref{SI-0}). For this reason, it is directly excluded from the counting.}
. 
However, only a few of them correspond to a set of $n$ {\it independent operators}. 
Consider that you have chosen $i$ independent operators, $\{\sigma_1, \dots, \sigma_i\}$, in $\Omega_n\backslash \{1\}$: with these operators you can generate a subset of $2^i$ operators of $\Omega_n$.
The number of operators left in $\Omega_n$ that are independent of the family $\{\sigma_1, \dots, \sigma_i\}$ is thus $|\Omega_n|-2^i=2^n-2^i$, which corresponds to the number of  possibilities for choosing another independent operator $\sigma_{i+1}$.
As a consequence, the number of different ways to sample $n$ independent operators from $\Omega_n$, i.e., the total number of GTs, is
\begin{equation}    
    \mathcal{N}_{GT} (n) 
        = \prod_{i=0}^{n-1} \, (2^n-2^i) 
        = \left(2^n\right)^n \, \prod_{i=1}^n \left(1-\frac{1}{2^i}\right)\,.
        \label{numerotrasfgauge}
\end{equation}
In this equation, one can recognise the q-Pochhammer symbol, $\left(\frac{1}{2},\frac{1}{2}\right)_n = \prod_{i=0}^{n-1} \big(1 - (\frac{1}{2})^{i+1}\big)$, 
a (strictly) decreasing function of $n$, converging rapidly ($n>5$) to its asymptotic value known as the Euler $\phi$-function, $\phi \big(\frac{1}{2}\big) \simeq 0.2887880950$.
For example, $\mathcal{N}_{GT}(3)=168$ and $\mathcal{N}_{GT}(4)=20160$; for $n>5$, the number of gauge transformations grows as $\mathcal{N}_{GT}(n)\sim 0.289 \times (2^n)^n$.  
Finally, the probability of getting a GT by 
drawing at random $n$ operators $\{\sigma_1, \dots, \sigma_n\}$ of~$\Omega_n$ converges asymptotically to a non-zero constant: 
\begin{align}
    \mathcal{P}_{GT} 
        = \frac
            {\left(\frac{1}{2}, \frac{1}{2}\right)_n}
            {\prod_{i=1}^n \left(1-\frac{i}{2^n}\right)} \;
        \underset{n \to\infty}{\longrightarrow} \; 
        \phi \left(\frac{1}{2}\right) \simeq 0.2887880950\,.
\end{align}
\section{Partition function of a spin model $\mathcal{M}$} \label{suppZ}
\subsection{Partition function and \textit{loops} of $\mathcal{M}$} In order to compute the complexity $c_{\mathcal{M}}$ of a 
model $\mathcal{M}$ from Eq.~(\EqFIMm) of the main text, one has first to compute the Fisher Information matrix $\mathbb{J}(\vecg)$,  
and, by extension, the partition function $Z_{\mathcal{M}}(\vecg)$ given in Eq.~(\ref{eq:graph_modelA}).
As each operator $\phi^\mu(\spin)$ only takes values in $\{-1, 1\}$, 
the exponential terms in Eq.~(\ref{eq:graph_modelA}) 
can be expanded as~\citepalias{*landau2013statistical,kramers1941statistics}
\[
    e^{g^\mu\phi^\mu(\spin)}
    =\cosh(g^\mu)+\phi^\mu(\spin)\sinh(g^\mu)
    =\cosh(g^\mu)\left[1+\phi^\mu(\spin)\tanh(g^\mu)\right]\,,
\]
which successively leads to the expressions for the partition function:
\begin{align}
    Z_{\mathcal{M}}(\vecg)
        &=
        \Bigg(\prod_{\mu\in \mathcal{M}}\cosh(g^\mu)\Bigg)
        \sum_{\spin\in\mathcal{S}} \;\prod_{\mu\in \mathcal{M}} \left[1+\phi^\mu(\spin)\tanh(g^\mu)\right]\,,\nonumber\\
        &= 
        \Bigg(\prod_{\mu\in \mathcal{M}}\cosh(g^\mu)\Bigg)
        \sum_{\spin\in\mathcal{S}} \; \Bigg[\sum_{\mathcal{M}'\subseteq\mathcal{M}}\prod_{\mu\in \mathcal{M}'}\phi^\mu(\spin)\tanh(g^\mu)\Bigg]\,,\nonumber
\end{align}
where the sum over $\mathcal{M}'\subseteq\mathcal{M}$ runs over all possible \emph{sub-models} (i.e. subsets) of 
$\mathcal{M}$ 
and the product is then taken over every operator of the sub-model $\mathcal{M}'$. 
The "empty model" $\mathcal{M}' = \{\emptyset\}$, with no interactions, is also included in the sum, 
considering that $\prod_{\mu\in\{\emptyset\}} \phi^\mu \tanh(g^\mu)= 1$.
In order to compute the sum over all configurations $\mathcal{S}$, one can exploit Eqs.~(\ref{eq:sum_phi}) and ~(\ref{eq:prod_op}), that lead to:
\begin{align}
\nonumber
\sum_{\spin\in\mathcal{S}} \prod_{\mu\in \mathcal{M}'}\phi^\mu(\spin)\tanh(g^\mu) 
&= \prod_{\mu\in \mathcal{M}'}\tanh(g^\mu) \sum_{\spin\in\mathcal{S}} \phi^{\oplus_{\mu \in \mathcal{M}'}}(\spin) \\
\nonumber
&= 2^n \prod_{\mu\in \mathcal{M}'}\tanh(g^\mu) \delta_{ \oplus_{\mu\in\mathcal{M'}} ,0} \, ,
\end{align}
where $\oplus_{\mu\in\mathcal{M'}}$ denotes the bitwise XOR operation between all the operators $\mu \in \mathcal{M}'$.
Here, the key observation is that $\oplus_{\mu\in\mathcal{M'}}=0$ if and only if
each spin occurs an even number of times (or none) among the operators of $\mathcal{M}'$. 
In this latter case, the operators of $\mathcal{M}'$ form a {\it loop}, 
such that $\prod_{\mu\in \mathcal{M}'}\phi^\mu(\spin)=1$ is equal to the identity operator.  
Let us name $\ell$ any sub-model $\mathcal{M}'$ that forms a {\it loop} 
and call $\mathcal{L}$ the set of all the loops $\ell$ of a given model $\mathcal{M}$ (including the empty loop $\{\emptyset\}$), allowing us to obtain
the expression in Eq.~(\Eqpartf) of the main text. The expansion of the partition function in loops is in the same spirit of cluster expansions methods in statistical physics (for a review see~\cite{pelizzola2005cluster}).

\subsection{Invariance of $Z_{\mathcal{M}}$ under gauge transformation}
In Eq.~(\Eqpartf) of the main text, the structure of the partition function depends only on few characteristics of the model~$\mathcal{M}$:
\begin{enumerate} 
    \item[{\em i)}]~the total number of operators $|\mathcal{M}|$, as they all appear in the product $\prod_{\mu\in \mathcal{M}}\cosh(g^\mu)$;  
    \item[{\em ii)}]~the structure of its set of {\it loops} $\mathcal{L}$:
    the number $|\mathcal{L}|$ of loops in the model (through the sum over $\mathcal{L}$); 
    the number $|\ell|$ of operators involved in each loop, named the {\it length of the loop} 
    (through the product over each operator $\mu$ of $\ell$); and finally which operators are involved in each loop. 
\end{enumerate} 
These properties are invariant under GTs, such that the structure of the partition function in Eq.~(\Eqpartf) of the main text remains invariant as well.
Indeed, consider two models, $\mathcal{M}$ and $\mathcal{M}^\prime=\mathcal{T}[\mathcal{M}]$, that are images of one another via a GT $\mathcal{T}$.
They verify the following properties:
\begin{itemize}
\item[{\em i)}] the two models have the same number of operators: $|\mathcal{M}|=|\mathcal{M}^\prime|$.
Indeed, we define the image of the set $\mathcal{M}$ by $\mathcal{T}$ as, $\mathcal{M'}=\mathcal{T}[\mathcal{M}] = \{\mathcal{T}[\phi^\mu], \mu\in\mathcal{M}\}$, and $\mathcal{T}$ is a bijection on the set of operators $\Omega_n$ (such that for all $\phi^\mu$, $\mathcal{T}[\phi^\mu]\in\Omega_n$ and if $\phi^\mu\neq\phi^\nu$ then $\mathcal{T}[\phi^\mu]\neq\mathcal{T}[\phi^\nu]$).
\item[{\em ii)}] the two models have the same loop structure. Indeed, if $\ell\in \mathcal{L}$ is a loop of the model $\mathcal{M}$, i.e. 
\[
\prod_{\mu\in\ell}\phi^\mu(\spin)=\phi^{\oplus_{\mu\in\ell}}=1,
\]
then $\ell'=\mathcal{T}(\ell)$ has the same length than $\ell$ and is a loop of the model $\mathcal{M}^\prime=\mathcal{T}[\mathcal{M}]$: 
\[
\prod_{\mu\in\ell^\prime}\phi^\mu(\spin)=\prod_{\mu\in\ell}\mathcal{T}[\phi^\mu(\spin)]=\mathcal{T}[\phi^0(\spin)]=1,
\]
where we used that $\mathcal{T}[\phi^\mu(\spin)\cdot\phi^{\mu'}(\spin)]=\mathcal{T}[\phi^\mu(\spin)]\cdot\mathcal{T}[\phi^{\mu'}(\spin)]$
(as $\mathcal{T}$ is an homomorphism of $\Omega_n$),
and that the identity element $\phi^0$ of $\Omega_n$ is invariant under $\mathcal{T}$.
Reciprocally, as $\mathcal{T}$ is a bijection, if $\ell'$ is a loop of $\mathcal{M}'$ then $\ell=\mathcal{T}^{-1}[\ell']$ is a loop of $\mathcal{M}$.
Finally, if $\ell_1$ and $\ell_2$ are two distinct loops of $\mathcal{M}$, then their respective images by $\mathcal{T}$ are two distinct loops of $\mathcal{M}'$. \\
In other words, if $\mathcal{L}$ is the group of loops of the model $\mathcal{M}$, then the group of loops associated to the model $\mathcal{M}'=\mathcal{T}[\mathcal{M}]$ is $\mathcal{L}'=\mathcal{T}[\mathcal{L}]$.
\end{itemize}

As a consequence, if two models are related by a GT, $\mathcal{M}^\prime=\mathcal{T}[\mathcal{M}]$, then they have the same value of 
complexity $c_{\mathcal{M}}=c_{\mathcal{M}^\prime}$. Indeed, the function under the integral in Eq.~(\EqFIMm) of the main text stays 
invariant under the change of variables from the model $\mathcal{M}$ to the model $\mathcal{M}'$.
Finally, gauge transformations define an equivalence relation between models, for which the structure (previously described, see {\it i)} and {\it ii)}) and the complexity $c_{\mathcal{M}}$ are invariant.
Gauge transformations thus allow us to partition all models into equivalence classes, that we call {\it complexity classes}. 
For instance, Fig.~\FigOne~of the main text displays several models for $n=4$ that belong to the same {\it complexity class} (highlighted in bold 
font in table~\ref{Table:n4}) for which $c_{\mathcal{M}}\simeq 2.8$. Note that, conversely, $c_{\mathcal{M}}=c_{\mathcal{M}^\prime}$ does not imply 
that $\mathcal{M}$ and $\mathcal{M}^\prime$ belong to the same complexity class. For example, the models $\mathcal{M}=\{s_1,s_2\}$ and 
$\mathcal{M}^\prime=\{s_1,s_2,s_1 s_2\}$ have both $c_{\mathcal{M}}=c_{\mathcal{M}^\prime}=2\log\pi$ (see Table~\ref{Table:n3}), but 
their structures are clearly different.\\

\section{Complexity classes and loop structure of spin models}\label{SI-3}
Let us highlight several interesting properties of models belonging to the same class of complexity. First, the {\it number of independent operators} 
$n_{\mathcal{M}}$ in a model $\mathcal{M}$ is invariant under GT and is thus a property of each complexity class (see~\ref{app:n_M}).
It can besides be (strictly) smaller than $n$: an important consequence is that any model with $n_{\mathcal{M}}<n$ is equivalent to a model involving 
only $n_{\mathcal{M}}$ spins. 
Second, the set of loops $\mathcal{L}$ of a model (including the empty loop~$\{\emptyset\}$) has the structure of a finite group, from which we show that the total number of loops of a given model $\mathcal{M}$ is of the form (see~\ref{app:L_structure}):
\begin{align}
    \label{eq:lambda_M}
    |\mathcal{L}|=2^{\lambda} \qquad
    {\rm with}\;\;\lambda = |\mathcal{M}|-n_{\mathcal{M}}\,.
\end{align}
The structure of the group $\mathcal{L}$ is an invariant of the class. 
Table~\ref{Table:n3} gives, for instance, a description of the loop structure for each class of complexity of models with $n=3$.
Finally, GTs also imply a duality relation between complexity classes of {\it complementary models}: 
each class of models with $|\mathcal{M}|$ operators corresponds to a {\it complementary class} of models with $(2^n-1)-|\mathcal{M}|$ operators 
that contains the same number of models (see~\ref{app:complementary_models} and Table~\ref{Table:n3}).

\subsection{Number of independent operators}
\label{app:n_M}
\begin{figure}[ht]
\centering
\includegraphics[width=0.48\columnwidth]{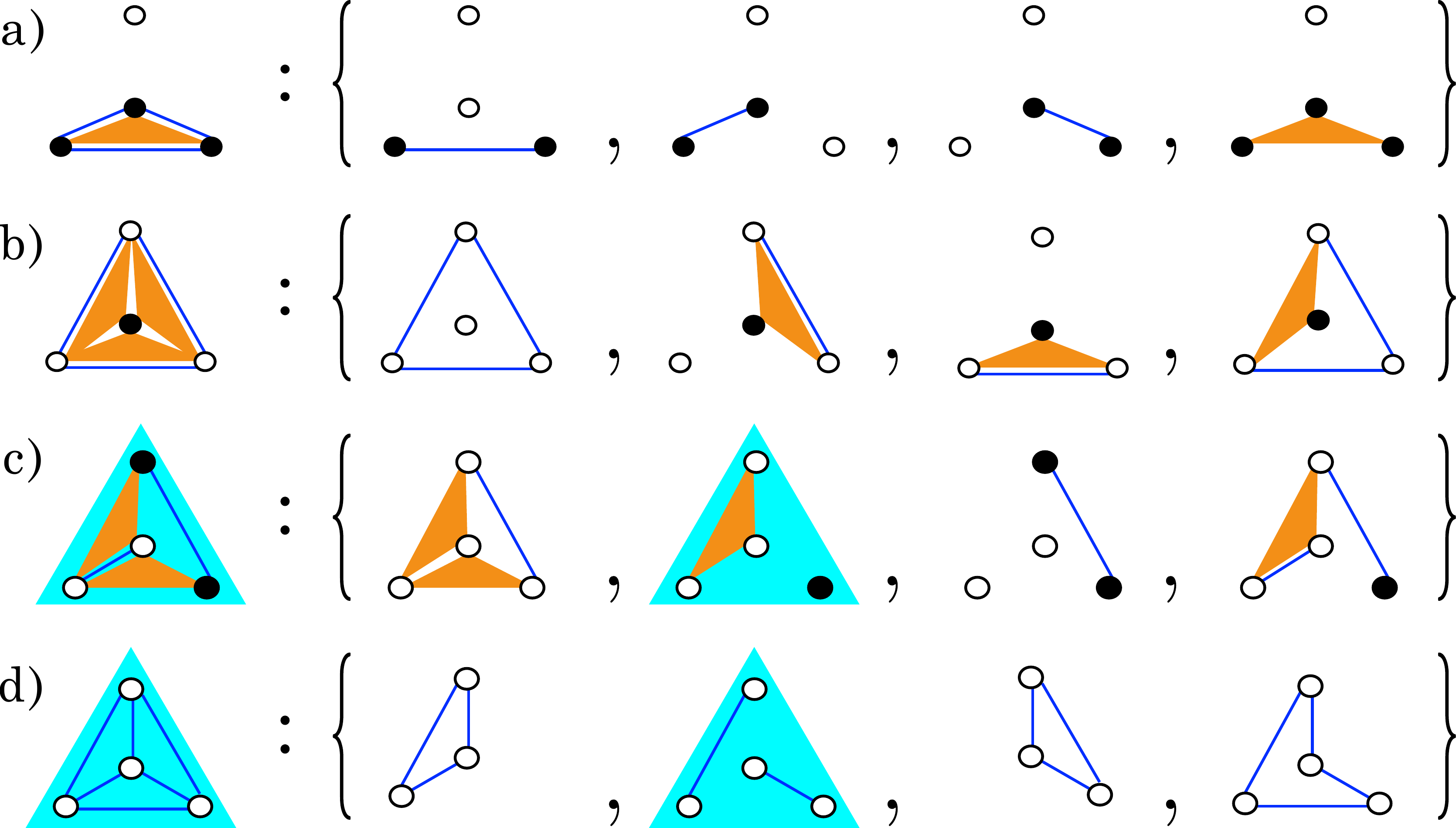}
\hspace{4.5mm}
\includegraphics[width=0.48\columnwidth]{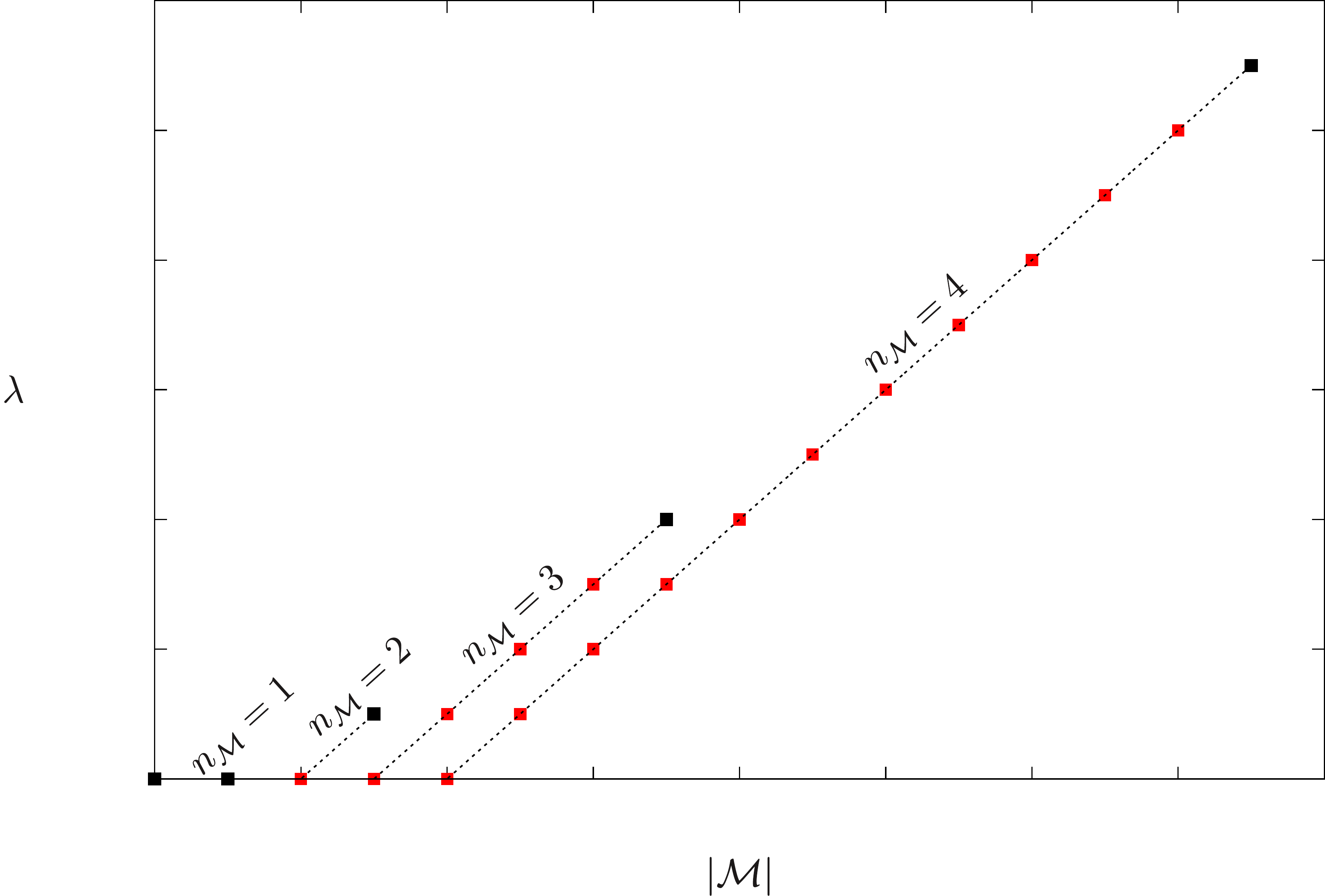}
\vspace{-2mm}
\caption{\label{fig:lambda} \label{Fig_k} \label{fig2} 
{\bf Left.} Decomposition into a generating set of loops for the models of Fig.~\FigOne~.
As these models belong to the same class, their (respective) sets of loops 
have the same structure and the same cardinality $2^\lambda$, where $\lambda=4$ is the number of generators (as shown here).
For model a), one can easily check that the $4$~loops of the set are independent,
as each of them contains at least one operator that doesn't appear
in the other $3$~loops (see \suppLoop).
Then, generating loops in each column (on the r.h.s.) are related by the same gauge transformation morphing the models into one another on the l.h.s. (see Fig.~\FigOne~). This shows that the loop structure of these models is the same. Any loop of a model can finally be obtained by combining a subset of its generating loops.
Note that the choice of the generating loops is not unique.
{\bf Right.} Dependence of $\lambda$, defined in \ref{app:L_structure}, as a function of the number of parameters $|\mathcal{M}|$ for different values of $n_\mathcal{M}$. 
Each value of $\lambda$ (squares) is extracted from the classification of all the possible models for $n=4$ (see table~\ref{Table:n4} in \ref{SI-6}). The dash lines corresponds to \eqref{eq:lambda_M} for different values of $n_{\mathcal{M}}$; the black squares, to the value of $\lambda_{\rm max}$ given in ~\eqref{eq:lambda_max}. 
Models with $\lambda=0$ are models with only independent operators, whereas models with $\lambda=\lambda_{\rm max}$ are equivalent to {\it sub-complete} models, i.e. complete models on a subset $n_{\mathcal{M}}$ of the $n$ spins.
}
\end{figure}

We define $n_{\mathcal{M}}$ as the maximum number of independent operators of a model $\mathcal{M}$, 
i.e. the maximum number of operators that can be taken in $\mathcal{M}$ without forming any loop.
Necessarily $n_{\mathcal{M}}\leq n$, because all operators can be generated by products of the $n$ spins.
Furthermore, the number of spin operators~\footnote{excluding the identity operator that doesn't correspond to any interaction, and thus doesn't belong to any model.} generated by $n_{\mathcal{M}}$ independent operators
is $2^{n_{\mathcal{M}}}-1$, which implies the following relations between $|\mathcal{M}|$ and $n_{\mathcal{M}}$ (see Fig.~\ref{fig:lambda} Right):
\begin{align}\label{app:eq:n_M}
    n_{\mathcal{M}} \leq |\mathcal{M}| \leq 2^{n_{\mathcal{M}}}-1 \qquad \Longleftrightarrow
    \qquad
    \log_2(|\mathcal{M}|+1)\leq n_{\mathcal{M}} \leq \min (n, |\mathcal{M}|)
    \,.
\end{align}
By definition, $n_{\mathcal{M}}$ is invariant under GT and is thus a property of each complexity class. Table~\ref{Table:n3}, for instance, reports the value of 
$n_{\mathcal{M}}$ for each class of models with $n=3$.
As an important consequence, any model with $n_{\mathcal{M}}<n$ 
can be mapped (through a GT) to a model involving 
only $n_{\mathcal{M}}$ spins.
See for instance the class displayed in Fig.~\FigOne~ of the main text: 
each model involves $|\mathcal{M}|=7$ operators for $n=4$ spins, but only $n_\mathcal{M}=3$ are independent;
consequently any of these models can be mapped through a GT to a model that involves only $3$ spins (see model a) in Fig.~\FigOne~).

\subsection{Loop structure of a model}\label{app:L_structure} 
For any model $\mathcal{M}$, the corresponding set of loops $\mathcal{L}$ has a finite cardinality\footnote{
Indeed, $\mathcal{L}$ is a subset of the set of all the partitions 
of $\mathcal{M}$, which has a finite cardinality (given by the Bell number $B_{|\mathcal{M}|}$).
}. 
Let us define the {\it disjunctive union} (or {\it symmetric difference}) of two loops $\ell_1$ and $\ell_2$ as the set 
that contains the operators that occur in $\ell_1$ but not in $\ell_2$ and {\em viceversa}: $\ell_1\oplus\ell_2 = (\ell_1\cup \ell_2)\backslash(\ell_1\cap \ell_2)$.
%
The set $\mathcal{L}$ is closed under disjunctive union $\oplus$: indeed if $\ell_1$ and $\ell_2$ are two elements of $\mathcal{L}$, then $\ell_1\oplus\ell_2$ is also in $\mathcal{L}$~\footnote{For example, consider the model $\mathcal{M}=\{s_1, s_2, s_3, s_1s_2, s_1s_3\}$: $\ell_1=\{s_1, s_2,s_1s_2\}$ and $\ell_2=\{s_1, s_3,s_1s_3\}$ are two loops of $\mathcal{M}$; $\ell_3=\ell_1\oplus\ell_2 = \{s_2, s_3, s_1s_2, s_1s_3\}$ is also a loop of $\mathcal{M}$. Note also that $\ell_2=\ell_1\oplus\ell_3$ and $\ell_1=\ell_2\oplus\ell_3$.}. We can thus find a {\it minimal generating set} of loops, of cardinality $\lambda\leq|\mathcal{L}|$, 
that can generates the whole set $\mathcal{L}$.
Finally, we note that the operation $\oplus$ is commutative and that any element of $\mathcal{L}$ is of order 2: for any loop $\ell$, $\ell\oplus\ell=\{\emptyset\}$. This way, the total number of loops that can be formed with $\lambda$ generating loops is $|\mathcal{L}|=2^{\lambda}$, including the empty loop $\{\emptyset\}$, and consequently, the total number of non-empty loops of any model is of the form $2^{\lambda}-1$.\\

Mathematically, for any model $\mathcal{M}$, the corresponding set of loop $\mathcal{L}$ forms a finite Abelian group 
associated with the operator of disjunctive union $\oplus$, the neutral element being the empty loop $\{\emptyset\}$. Each element of this group is of order two, which implies that the cardinality of the group is of the form $2^\lambda$, where $\lambda$ is then the cardinality of the minimal set of generators of~$\mathcal{L}$.\\

Let us now prove the relation in ~\eqref{eq:lambda_M}.
Consider the model $\mathcal{M}$ with $|\mathcal{M}|$ operators, of which at most 
$n_\mathcal{M}$ are independent. Let us take one maximal subset of independent operators of $\mathcal{M}$ and call it $I_{\mathcal{M}}$: by construction, the number of operators in this subset is $|I_{\mathcal{M}}|=n_{\mathcal{M}}$ and the set of loops that can be formed with these operators is necessarily empty.
If $|\mathcal{M}|=n_\mathcal{M}$, then the set of loops of $\mathcal{M}$ is empty,
which is consistent with $\lambda=0$ in ~\eqref{eq:lambda_M}.
If instead $|\mathcal{M}| > n_\mathcal{M}$, then, for any other operators $\phi^\nu\in\mathcal{M}\backslash I_{\mathcal{M}}$, 
we can find a subset $\mathcal{P}^\nu\subseteq I_{\mathcal{M}}$ 
such that the set $\ell^\nu=\{\phi^\nu\}\cup\mathcal{P}^\nu$ is a loop of $\mathcal{M}$, i.e.
\begin{align}
    \phi^\nu \times\Bigg(\prod_{\mu\in\mathcal{P}^\nu} \phi^\mu \Bigg) = 1\,.
\end{align}
This procedure, for each different operator $\phi^\nu\in\mathcal{M}\backslash I_{\mathcal{M}}$, produces a different loop, so the set of  $|\mathcal{M}|-n_\mathcal{M}$ loops  
$\ell^\nu=\{\phi^\nu\}\cup\mathcal{P}^\nu$ built in this way is a minimal generating set of $\mathcal{L}_{\mathcal{M}}$. Note indeed that each loop $\ell\in\mathcal{L}_{\mathcal{M}}$ can be decomposed in the loops $\ell^\nu$ in an unique way. 
Therefore $\lambda=|\mathcal{M}|-n_\mathcal{M}$.
Table~\ref{Table:n3}, reports the values of $|\mathcal{M}|$, $n_{\mathcal{M}}$ and $\lambda$ for $n\le 4$ and  ~\eqref{eq:lambda_M} can be verified for each class of complexity. For a fixed number $n_{\mathcal{M}}$ of independent operators, $\lambda$ thus grows linearly as a function of the number of parameters of $\mathcal{M}$  up to a maximum value (see Fig.~\ref{fig:lambda}),
\begin{align}\label{eq:lambda_max}
    \lambda_{\rm max} = |\Omega_{n_{\mathcal{M}}}\backslash\{1\}|-n_{\mathcal{M}} = 2^{n_{\mathcal{M}}}-1-n_{\mathcal{M}}\,,
\end{align}
where $\Omega_{n_{\mathcal{M}}}$ is the set of operators that can be generated from $n_{\mathcal{M}}$ spins. The value of $\lambda_{\rm max}$ is associated with the maximum set of loops that can be generated from $n_{\mathcal{M}}$ spins, that 
corresponds to the class of complexity of {\it sub-complete models} on $n_{\mathcal{M}}$ spins, i.e. complete models on a subset $n_{\mathcal{M}}$ of the $n$ spins.

\subsection{Complexity classes of complementary models} 
\label{app:complementary_models}
Consider a model $\mathcal{M}$ with $|\mathcal{M}|$ operators, we define the {\it complementary model} $\mathcal{M}_c$ as the unique model that contains all the operators that are not in $\mathcal{M}$, i.e. $\mathcal{M}_c = \Omega_n \backslash \{1\} \backslash\mathcal{M}$, which can also be written as:
\begin{align}
    \mathcal{M}_c \cup \mathcal{M} = \Omega_n\backslash\{1\}\qquad {\rm and} \qquad
    \mathcal{M}_c \cap \mathcal{M} = \emptyset\,.
\end{align}
By definition, $\mathcal{M}_c$ has exactly $|\mathcal{M}_c| = 2^n - 1 - |\mathcal{M}|$ operators. 
Using the properties of gauge transformations, it can be shown that, if two models belong to the same complexity class~$\mathcal{C}$, then their respective complementary models also belong to the same class $\mathcal{C}_c$ (see proof below). 
As a consequence, the two corresponding classes of complexity (named {\it complementary classes}) have the same cardinality and the number of complexity classes with $|\mathcal{M}|$ parameters is equal to the number of classes with $2^n-1-|\mathcal{M}|$ parameters.
Observe, for instance, in Table~\ref{Table:n3} the symmetry in the cardinality of classes of complementary models (starting from the first line and the last line of the table).\\

{\it Proof:} Consider a model $\mathcal{M}$, its complementary model $\mathcal{M}_c$, and their respective classes of complexity, $\mathcal{C}$ and $\mathcal{C}_c$. Let us take a GT $\mathcal{T}$ and define the transformed models $\mathcal{M}^\prime = \mathcal{T}[\mathcal{M}]$ and $\mathcal{M}^\prime_c = \mathcal{T}[\mathcal{M}_c]$. By definition, $\mathcal{M}^\prime\in\mathcal{C}$ and  $\mathcal{M}^\prime_c\in\mathcal{C}_c$. 
As GT are bijections of the space of operators and $\mathcal{M}_c \cap \mathcal{M} = \emptyset$, the new sets of operators obtained after GT are necessarily disjoint: $\mathcal{M}^\prime_c \cap \mathcal{M}^\prime = \emptyset$.
Besides, $|\mathcal{M}_c^\prime|+|\mathcal{M}^\prime| = |\mathcal{M}_c|+|\mathcal{M}|= |\Omega_n\backslash\{1\}|$, 
such that the two models also verify that $\mathcal{M}^\prime_c \cup \mathcal{M}^\prime = \Omega_n\backslash\{1\}$. In other words, $\mathcal{M}^\prime_c$ is the complementary model of $\mathcal{M}^\prime$. We thus obtain that, if two models belong to the same class of complexity $\mathcal{C}$, then their respective complementary models also belong to the same class $\mathcal{C}_c$, thus called {\it complementary class} of $\mathcal{C}$. 

\section{A general argument for the calculation of $c_{\mathcal{M}}$}\label{SI-4}
In this section we provide a general argument for the calculation of the complexity  $c_{\mathcal{M}}$: 
in the first part we compute the complexity of models with only independent operators and show that any independent operator (that doesn't enter in a loop) contributes as $\log\pi$ to $c_{\mathcal{M}}$;
while in the second part we suggest that this value constitutes an upper bound for the complexity of any operator. 

\subsection{Models with only independent operators}\label{SI-4.1}
At fixed number of spins $n$, every model with $|\mathcal{M}|=n_{\mathcal{M}}$ independent operators and non-degenerated parameters belongs to the same class of complexity (see SI-3), which is the class of models with 
$\mathcal{L}=\{\{\emptyset\}\}$. The number of models in such a class is 
\begin{align}\label{eq:indep_param:N_ind}
    \mathcal{N}^{\rm ind}_n(n_{\mathcal{M}}) = \frac{\prod_{i=0}^{n_{\mathcal{M}}-1} (2^n-2^i)}{n_{\mathcal{M}} !}\,,
\end{align}
which is the number of possible ways~\footnote{
obtained with the same argument than the number of gauge transformations in \ref{SI-1}, see Eq.~(\ref{numerotrasfgauge}).
}
of choosing $n_{\mathcal{M}}$ independent operators in $\Omega_n\backslash\{1\}$
divided by their permutations. Notice that $1\leq n_{\mathcal{M}}\leq n$, as $n+1$ operators are necessarily forming at least one loop. 
In particular, for models with $|\mathcal{M}|=1$ operator, we recover that $\mathcal{N}^{\rm ind}_n(1)=\vert{\Omega_n}\vert$ (any operator can be chosen); and, for models with $n_{\mathcal{M}}=n$ independent operators, we obtain that $\mathcal{N}^{\rm ind}_n(n)=\mathcal{N}_{GT}/n!\,$.
For instance, with $n=4$, Eq.~(\ref{eq:indep_param:N_ind}) gives: $\mathcal{N}^{\rm ind}_4(2)=105$ for $n_{\mathcal{M}}=2$, $\mathcal{N}^{\rm ind}_4(3)=420$ 
and $\mathcal{N}^{\rm ind}(4)=840$, 
that match with the results in Table~\ref{Table:n4}.\\

The partition function of a model with $n_\mathcal{M}=\vert \mathcal{M} \vert$ independent operators contains 
only the first term of Eq.~(\Eqpartf) of the main text, $Z(\vecg)= 2^n \prod_{\mu \in \mathcal{M}}\cosh(g^\mu)$. 
The Fisher information matrix (FIM), in Eq.~(\EqFIMm) of the main text, is therefore diagonal and reads $J^{\rm ind}_{\mu \nu}(\vecg) = \delta_{\mu \nu}[1-\tanh^2(g^\mu)]$, which finally leads to the complexity term:
\begin{align}
\label{eq:cM:indep}
    e^{c_{\mathcal{M}}} 
        = \int_{\R^{\vert \mathcal{M} \vert}} \prod_{\mu \in \mathcal{M}} \sqrt{ 1-\tanh^2(g{\mu}) }\;\dd \vecg 
        = \left[\int_{\R} \sqrt{1-\tanh^2(g)}\;\dd g \right]^{\vert \mathcal{M} \vert}
        = \pi^{\vert \mathcal{M} \vert}\,.
\end{align}
For the same reason, any operator of a model, that doesn't enter in any loop of the model, contributes with a  
term  $\log\pi$ to the complexity $c_{\mathcal{M}}$. 
Indeed, let us consider a model $\mathcal{M}$ with $\vert \mathcal{M}\vert$ operators, including $K$ independent $\phi^1, \dots\phi^K$. We can then introduce a model $\mathcal{M}^\prime$ formed by the set of non-independent operators of $\mathcal{M}$, and a model $\mathcal{M}_{\rm ind}=\{\phi^1, \dots\phi^K\}$, so that $\mathcal{M}=\mathcal{M}^\prime\cup\mathcal{M}_{\rm ind}$, $|\mathcal{M}|=|\mathcal{M}^\prime|+K$ and $\mathcal{L}_{\mathcal{M}^\prime}=\mathcal{L}_{\mathcal{M}}$. 
As a consequence the partition function in Eq.~(\Eqpartf) of the main text
can be factorized in the two models: 
\begin{align}
    Z_{\mathcal{M}}(\vecg) 
        &= 2^n \prod_{\mu\in\mathcal{M}_{\rm ind}}\cosh(g^\mu)
                \prod_{\mu\in\mathcal{M}\prime}\cosh(g^\mu) 
        \left[\sum_{\mathcal{\ell}\in \mathcal{L_{\mathcal{M}^\prime}}}\prod_{\mu\in\mathcal{\ell}}\tanh (g^\mu)\right]\,\\
        &=\frac{1}{2^n}\, Z_{\mathcal{M}_{\rm ind}}(g^1, \dots, g^K) \,Z_{\mathcal{M}^\prime}(g^{K+1}, \dots, g^{\mathcal{M}})
\end{align}
Using the previous argument, we obtain that the FIM of $\mathcal{M}$ is a block matrix:
\begin{align}
    \mathbb{J}_{\mathcal{M}}(\vecg)=
        \begin{pmatrix}
        \mathbb{J}^{\rm ind}(g^{\,[1,K]}) & 0 \\
        0 &\mathbb{J}_{\mathcal{M}\prime}(\vecg^\prime) 
\end{pmatrix}
\end{align}
where $\mathbb{J}^{\rm ind}$ is the diagonal matrix previously introduced (for models with only independent operators). Finally, using the property of the determinant, $\det \mathbb{J} = \det[\mathbb{J}^{\rm ind}]\det [\mathbb{J}_{\mathcal{M}\prime}]$, 
we obtain that
\begin{align}
    e^{c_{\mathcal{M}}}
        = e^{c_{\mathcal{M}_{\rm ind}}}\;\times\;e^{c_{\mathcal{M}^\prime}} 
        = \pi^K\;\times\;e^{c_{\mathcal{M}^\prime}}
    \,.
\end{align}

Notice that, in ~\eqref{eq:cM:indep}, the stochastic complexity for $|\mathcal{M}|$ independent operators is linear in the number 
of operators $|\mathcal{M}|$, which is of the same form than 
the first penalty term in the BIC (penalty due to the number of operators -- see Eq.~\EqMDLc~of the main text).

\subsection{General argument} \label{SM-4.2}
Given a dataset $\hspin=(\spin^{(1)},...,\spin^{(N)})$ where $\spin^{(i)}$ are $n$-spins configurations, the maximum likelihood of exponential models defined in ~\eqref{eq:graph_modelA}, is achieved when the  empirical averages $\empavg^\mu$ \eqref{eq:empirical_av} of the operators $\phi^{\mu}$ match the population averages $\ensavg^{\mu}$ \eqref{eq:ens_averages}, and takes the value:
\[
P(\hspin|\hvecg,\mathcal{M})=e^{-NS(\vecempavg)},
\]
where $S(\vecempavg)$ is the entropy of $P(\spin|\hvecg)$.
So we can introduce a delta function enforcing Eq. (\ref{eq:maxlikelihood}) in Eq. (\ref{mdlc}) for each operator $\mu\in \mathcal{M}$, resulting in
\begin{equation}
\label{mdl2}
\left(\frac{2\pi}{N}\right)^{|\mathcal{M}|/2}\sum_{\hat \spin} P(\hspin|\hvecg,\mathcal{M})=\left(\frac{2\pi}{N}\right)^{|\mathcal{M}|/2}\int_{-1}^1\! d\vecensavg e^{-NS(\vecensavg)}
\sum_{\hat \spin} \prod_{\mu\in M} \delta\left(\ensavg^\mu-\frac{1}{N}\sum_{i=1}^N\phi^\mu(\spin^{(i)})\right)\, .
\end{equation}
The sum over samples is zero unless $\vecensavg$ can be realized in at least one sample. This means that $\ensavg^\mu=(2k^\mu-N)/N$ can only attain $N+1$ values for $k^\mu=0,1,\ldots,N$. Yet not all values of $\vecensavg$ can be realized. For instance, if $\phi^1=s_1$, $\phi^2=s_2$ and $\phi^3=s_1s_2$, when $\empavg^1=\empavg^2=1$ there are no samples for which $\empavg^3\ne 1$. So let 
\[
\mathcal{F}=\left\{\vecensavg':~\exists\hspin~~(\ensavg')^\mu={\empavg^\mu}(\hspin)~\forall \mu\in \mathcal{M}\right\}
\]
be the set of feasible values of $\vecensavg$, known as the \textit{marginal polytope} \cite{wainwright2003variational}. Then the integral in Eq. (\ref{mdl2}) becomes a sum 
\begin{equation}
\label{ }
\left(\frac{2\pi}{N}\right)^{|\mathcal{M}|/2}\sum_{\hspin} P(\hspin|\hvecg,\mathcal{M})=
\left(\frac{2\pi}{N}\right)^{|\mathcal{M}|/2}
\sum_{\vecensavg\in \mathcal{F}}Q(\vecensavg) e^{-NS(\vecensavg)}
\end{equation}
where $Q(\vecensavg')$ is the number of samples $\hat \spin$ for which ${\empavg^\mu}(\hat \spin)=(\ensavg')^\mu$ for all $\mu\in \mathcal{M}$. $Q(\vecensavg)$ can be estimated in the weak dependence limit where it is given by
\[
Q(\vecensavg)\simeq\prod_{\mu\in \mathcal{M}}{N \choose N\frac{1+\ensavg^\mu}{2}}\simeq 
\left(\frac{2}{\pi N}\right)^{|\mathcal{M}|/2}
e^{NS(\vecensavg)}
\prod_{\mu\in \mathcal{M}}\left[1-\left(\ensavg^\mu\right)^2\right]^{-1/2}.
\]
So 
\begin{equation}
\label{ }
\left(\frac{2\pi}{N}\right)^{|\mathcal{M}|/2}\sum_{\hat \spin} P(\hat \spin|\hat \vecg, \mathcal{M})\simeq 
\left(\frac{2}{N}\right)^{|\mathcal{M}|}
\sum_{\vecensavg\in \mathcal{F}}\prod_{\mu\in \mathcal{M}}\left[1-\left(\ensavg^\mu\right)^2\right]^{-1/2}\simeq
\int_{\mathcal{F}}d\vecensavg \prod_{\mu\in \mathcal{M}} \left[1-\left(\ensavg^\mu\right)^2\right]^{-1/2}
\end{equation}
where we have turned the sum over $\ensavg^\mu$ into an integral, observing that $\textrm{d}\ensavg^\mu= 2/N$.
In the limit $N\to\infty$ we finally get 
\begin{equation}
\label{mdlcF}
e^{c_{\mathcal{M}}}=\int d\vecg \sqrt{\det J (\vecg)}\simeq\int_{\mathcal{F}}d\vecensavg \prod_{\mu\in \mathcal{M}} \left[1-\left(\ensavg^\mu\right)^2\right]^{-1/2}.
\end{equation}
This is a quite interesting result. It tells us that the complexity of a model is related to how operators of the model constrain the values that the expected values of other operators can take. All integrals in Eq. (\ref{mdlcF}) are over a subset $\mathcal{F}$ of the hypercube $[-1,1]^{|\mathcal{M}|}$, with the same integrand. Then the complexity uniquely depends on the volume of $\mathcal{F}$ under the measure $p(\vecensavg)\propto\prod_{\mu \in \mathcal{M}}(1-(\ensavg^{\mu})^2)^{-1/2}$.

The approximation used becomes exact in the case of independent operators and is likely an upper bound otherwise. 

As a corollary, we find that the most complex models are those where all operators are independent and $\mathcal{F}=[-1,1]^{|\mathcal{M}|}$. In this case the integral takes the value $\pi^{|\mathcal{M}|}$ (see \ref{SI-4.1}). The least complex models, instead, are those where operators constrain themselves as much as possible. This correspond to models where all operators depend on the same subset of spins.

\section{The complexity of complete models}\label{SI-5}
In this section we derive an analytic expression for the complexity of {\it complete models} exploiting the invariance under reparametrization of Jeffreys prior \cite{jeffreys1946invariant} distribution over the parameters.
\subsection{Properties of the \textit{complete model}} 
The {\it complete model}~$\overline{\mathcal{M}}$ involves all the $2^n-1$ operators of $\Omega_n\backslash\{1\}$.
This model presents the peculiarity that the number of parameters $|\overline{\mathcal{M}}|$ equals the number of independent parameters that are needed to specify a generic distribution $p(\spin)$ on the spin configurations. In order to make this more explicit let us label by an integer $i\in\{0,1,..., 2^n-1\}$ all configurations $\spin^i\in \mathcal{S}$ and let $p^{i}= p(\spin^i)$. We can take $p^{0}=p(\spin^0)$ to be constrained by normalisation: $\sum_{i=0}^{2^n-1} p^{i}=1$, which leaves $2^n-1= |\overline{\mathcal{M}}|$ free parameters $p^i$, $i\in\{1,..., 2^n-1\}$. 
Re-writing (\ref{eq:graph_modelA}) 
with this notation,
\begin{align}
    p^{i} = \exp\left[\sum_{\mu=1}^{2^n-1} g{\mu}\,\phi^{\mu}(\spin^{i}) - \log Z_{\mathcal{M}}(\vecg)\right]\,,
    \label{GtoP}
\end{align}
taking the log, multiplying by $\phi^\nu(\spin^{i})$ and summing over $i$ with the relation in ~\eqref{orth}, leads to 
\begin{equation}
\label{PtoG}
g^\nu=\frac{1}{2^n} \sum_{i=0}^{2^n-1}\phi^\nu(\spin^i)\log p^{i}, \qquad {\rm where }\;\; p^{0} = 1-\sum_{i=1}^{2^n-1}p^{i}\,.
\end{equation}
We call this model {\it complete} in the sense that (\ref{GtoP}) and (\ref{PtoG}) define a bijection between the sets of parameters $\vecg=  \{g^{\mu}\}_{\mu\in\{1,...,2^n-1\}}$ and the sets of $2^n-1$ probabilities $\vecp = \{p^{i}\}_{i\in\{1,...,2^n-1\}}$, provided that one includes the values $\pm\infty$ as legitimate values for the individual parameters $g^\mu$.
This shows that this basis of models is {\it complete} in the sense that any probability distribution can be represented within this class of models.

\subsection{Complexity of the complete model} 
The bijection between parameters $\{g^\mu\}$ and probabilities $\{p^i\}$ indicates that $\vecp$ constitutes a suitable parametrization for the {\it complete model}. The Fisher Information matrix in $\vecp$ reads:
\begin{align}
    J_{ij}(\vecp) 
    = -\left\langle \partial_{p^i}\partial_{p^j} \log P(\spin\,|\,\vecp, \mathcal{M}) \right\rangle_{P}
    = \frac{\delta_{i,j}}{p^{i}} +\frac{1}{p^0}\,,
    \qquad\qquad {\rm for\;\; all }\; 
    (i, j)\in\{1,...,2^n-1\}^{\,2}\,.
\end{align}
Using that the {\it volume element} 
\begin{align}
    \dd V = \sqrt{\det \mathbb{J}(\vecg)}\;\dd \vecg = \sqrt{\det \mathbb{J}(\vecp)}\;\dd \vecp\,,
\end{align}
is invariant under re-parametrisation~\cite{Amari}, we can express the complexity 
in the new set of parameters $\vecp$:
\begin{align}
    c_{\mathcal{M}} = \log \int \dd \vecp\;\sqrt{\det \mathbb{J}(\vecp)}\,.
\end{align}
In the $p$-parameters, the determinant of $\mathbb{J}(\vecp)$ can be more easily worked out by rewriting the FIM as:
\begin{equation}
\mathbb{J}(\vecp)=\mathbb{D}\left(\matId+\mathbf{v}\mathbf{w}^t\right) 
\end{equation}
where $\mathbb{D}$ is a diagonal matrix with entries $D_{ii}=1/p_i$, $\matId$  is the identity matrix, and, $\mathbf{v}$ and $\mathbf{w}$ are two vectors with elements
\begin{align}
v_i=p^i \quad \;\text{and}\;\quad w_j=\frac{1}{p^0}\,. 
\end{align}
Finally by using the properties of the determinant, $\det(\mathbb{D}\left( \matId+\mathbf{v}\mathbf{w}^t\right))=\det\mathbb{D}\,\det(\matId+\mathbf{v}\mathbf{w}^t)$ and $\det(\matId+\mathbf{v}\mathbf{w}^t)=1+\mathbf{v}^t \mathbf{w}$, one gets
\begin{equation}
\det \mathbb{J}(\vecp)=\prod_{i=0}^{2^n-1}\frac{1}{p^{i}}\,.
\end{equation}
So the complexity of the complete model is
\begin{equation}
\label{cMcompA}
c_{\overline{\mathcal{M}}}=\log\int_{[0,1]^{2^n}}\!\textrm{d}\vecp \, \delta\left(\sum_{i=0}^{2^n-1}p^{i}-1\right)\prod_{i=0}^{2^n-1} \frac{1}{\sqrt{p^{i}}}=2^{n-1}\log\pi-\log\Gamma(2^{n-1})
\end{equation}

\section{Complexity: numerical estimates}\label{SI-6}
In this Section we deal with the computation of the complexity penalty $c_{\mathcal{M}}$, defined in Eq.~(\EqFIMm) of the main text. We start by considering the generic model with an arbitrary loop structure and we derive the expression of the complexity integral for a model with a single loop. Finally we assess numerically the complexity for all the models on systems of $n\leq 4$ spins. Here we will focus on models with non-degenerated parameters, leaving the discussion on degeneracy to \ref{SI-7}.

\subsection{Generic model}\label{app:complex_genericND}\label{app:complex_general}
In this section we will derive an expression for the complexity integral of a generic model suitable for numerical integration. 
Since the contribution to the complexity of independent operators has been derived in \ref{SI-4}, here we will focus on models in which there are no independent operators, i.e. every operator participates in at least into one loop.

The elements of the FIM in Eq.~(\EqFIMm) of the main text, obtained by taking the derivatives of the logarithm of the partition function Eq.(\Eqpartf) of the main text, are for a generic model $\mathcal{M}$:
\begin{equation}
J_{\mu\nu} = \begin{cases}
  \Bigl(1 - \gamma_{\mu}^2\Bigl)\Bigl(1 - 2 \frac{\chi_{\mu}}{1 + \chi} - \frac{1 - \gamma_{\mu}^2}{\gamma_{\mu}^2} \frac{\chi_{\mu}^2}{(1+\chi)^2}\Bigl) &\text{for}\qquad \mu=\nu\\ 
 &\\
 \frac{1 - \gamma_{\mu}^2}{\gamma_{\mu} (1 + \chi)} \frac{1 - \gamma_{\nu}^2}{\gamma_{\nu} (1 + \chi)} \Bigl(\chi_{\mu,\nu} (1 + \chi) - \chi_{\mu} \chi_{\nu}\Bigl)&\text{for}\qquad \mu\ne \nu
\end{cases}
\label{eq:Fisher1}
\end{equation}
where we have defined
\begin{subequations}
\begin{eqnarray}
\gamma_{\mu} &=& \tanh g^{\mu}  \\
\chi &=& \sum_{\mathcal{\ell}\in \mathcal{L_{\mathcal{M}}\setminus\{\emptyset\}}}\, \prod_{{\mu} \in \mathcal{\ell}} \gamma_{\mu}\\
\chi_{\mu} &=&  \sum_{\mathcal{\ell} | \mu } \prod_{\nu \in \mathcal{\ell}} \gamma_{\nu} \\
\chi_{\nu,\mu} &=&  \sum_{\mathcal{\ell} | \{\nu,\mu\} } \prod_{\sigma \in \mathcal{\ell}} \gamma_{\sigma}
\end{eqnarray}
\label{eq:chis}
\end{subequations}
\noindent 
and "$\mathcal{\ell} | \mu $" ("$\mathcal{\ell} | {\nu,\mu} $") refers to the loops in $\mathcal{L_{\mathcal{M}}}$ in which $g^{\mu}$ ($g^{\mu}$ and $g^{\nu}$) enters.
In light of (\ref{eq:Fisher1}) the FIM can be expressed as

\begin{equation}
\mathbb{J}(\vecg)=\mathbb{A}(\vecg) + \mathbb{W}(\vecg)
\label{eq:Fisher2}
\end{equation}
where $\mathbb{A}(\vecg)$ is a diagonal matrix with $A_{\mu\mu}=J_{\mu\mu}-W_{\mu\mu}$ and $\mathbb{W}(\vecg)$ is defined as: 

\begin{equation}
\begin{split}
W_{\mu\nu} = \Bigl(\frac{1 - \gamma_{\mu}^2}{\gamma_{\mu}}\Bigl) \frac{\chi_{\mu}}{1 + \chi}\Bigl(\frac{1 - \gamma_{\nu}^2}{\gamma_{\nu}}\Bigl) \frac{\chi_{\nu}}{1 + \chi} \Bigl(\frac{\chi_{\mu,{\nu}}}{\chi_{\mu} \chi_{\nu}}(1  +\chi) - 1\Bigl)
\end{split}.
\end{equation}
Splitting $\mathbb{J}(\vecg)$ as in (\ref{eq:Fisher2}) allows us to rewrite the determinant of the FIM

\begin{equation}
\det \mathbb{J}(\vecg)=\det (\mathbb{A}(\vecg) + \mathbb{W}(\vecg)) = \det \mathbb{A}(\vecg) \det (\matId + \mathbb{A}^{-1}(\vecg) \mathbb{W}(\vecg))
\label{eq:detJ1}
\end{equation}
and to exploit the fact that $\det \mathbb{A}(\vecg)$ is simply the product of the diagonal entries of $\mathbb{A}$
\begin{equation}
\det \mathbb{A}(\vecg)=\prod_{\mu}\Bigl(1 - \gamma_{\mu}^2\Bigl)\Bigl(1 - \frac{1+\gamma_{\mu}^2}{\gamma_{\mu}^2} \frac{\chi_{\mu}}{1 + \chi}\Bigl)
\label{eq:detA}
\end{equation}
and that $\mathbb{A}^{-1}(\vecg)$ is a diagonal matrix with entries $A{^{-1}}_{\mu\mu}=(A_{\mu\mu})^{-1}$. Notice that $\matId$ in ~\eqref{eq:detJ1} is the identity matrix.
Substituting (\ref{eq:detA}) in (\ref{eq:detJ1}) one gets:
\begin{equation}
\begin{split}
\det \mathbb{J}(\vecg) = \prod_{\mu} \Bigl(1-\gamma_{\mu}^2\Bigl) \Biggl[\prod_{\mu} \Bigl( 1 - \frac{\chi_{\mu}}{1 + \chi}\frac{1+\gamma_{\mu}^2}{\gamma_{\mu}^2}\Bigl)\Biggl]\bigl( \det (\matId +\mathbb{B}(\vecg)) \bigl) \\
B_{\mu\nu} = \frac{\gamma_{\mu}}{\gamma_{\nu}} \frac{1-\gamma_{\nu}^2}{1 + \chi} \frac{\chi_{\mu,\nu}(1 + \chi) - \chi_{\mu} \chi_{\nu}}{(\gamma_{\mu}^2 (1 + \chi) - (1+\gamma_{\mu}^2)\chi_{\mu})}
\end{split}
\label{eq:detJ2}
\end{equation}
where we have defined the matrix $\mathbb{B}(\vecg)=\mathbb{A}^{-1}(\vecg)\mathbb{W}(\vecg)$.
Replacing the determinant of the FIM ( Eq.~(\ref{eq:detJ2})) in the complexity integral (Eq.~(\EqFIMm)~ in the main text) and performing the change of variables $g^{\mu}\rightarrow\gamma_{\mu}$, defined in ~\eqref{eq:chis}, yields

\begin{equation}
    \label{cM1}
  e^{c_\mathcal{M}}=\pi^{\vert \mathcal{M}\vert}\int_{\left[-1,1\right]^{\vert \mathcal{M}\vert} }\dd\boldsymbol{\gamma}\,q(\boldsymbol{\gamma})\Biggl[\prod_{\mu\in\mathcal{M}}\sqrt{\Bigl( 1 - \frac{\chi_{\mu}}{1 + \chi}\frac{1+\gamma_{\mu}^2}{\gamma_{\mu}^2}\Bigl)}\Biggl]\sqrt{ \det (\matId +\mathbb{B})}
\end{equation}
where 
\begin{equation}
q(\boldsymbol{\gamma})=\frac{1}{\pi^{\vert \mathcal{M}\vert}}\prod_{\mu} \Bigl(1-\gamma_{\mu}^2\Bigl)^{-1/2}\,. 
\end{equation}
Now (\ref{cM1}) is prone for standard Monte Carlo integration by random sampling $\boldsymbol{\gamma}$ according to the pdf $q(\boldsymbol{\gamma})$ on its bounded support. As one could easily check, $q(\boldsymbol{\gamma})$ is the measure induced by a set of $\vert\mathcal{M}\vert$ independent operators on the hypercube $\left[-1,1\right]^{\vert\mathcal{M}\vert}$ (see \ref{SI-4}).

\subsection{Models with a single loop} 
\label{App:SingleLoopND} 
We consider models with all their operators involved in a single loop.
The contribution to the complexity $c_{\mathcal{M}}$ of any supplementary independent operator (not involved in the loop) was studied in~\ref{SI-4} (contribution of $\log\pi$ for each supplementary operator), and will not be considered here.  
Single loop models $\mathcal{M}$ are such that $\mathcal{L}=\{\{\emptyset\}, \ell\}$, where $\ell=\{\phi^1, \dots, \phi^{\vert \mathcal{M}\vert}\}=\mathcal{M}$. All such models with fixed number of operators belong to the same class of complexity. The cardinality of this class can be assessed thinking at the single loop model with  $\vert \mathcal{M}\vert$ operators as a set of $\vert \mathcal{M}\vert-1$ independent operators plus an operator being the product of the independent ones. Following this reasoning the number of single loop models on a $n$ spins system is:

\begin{align}\label{eq:NumSingleLoop}
    \mathcal{N}_{\rm 1 \, loop} = \prod_{l=0}^{\vert \mathcal{M}\vert-2}\frac{(2^n-2^l)}{\vert \mathcal{M}\vert!}\, ,
\end{align}
where the previous formula was derived in full analogy with the number of models with only independent operators (see SI-4). 

\subsubsection*{\textbf{Complexity of single loop models}}\label{App:ComplSingleLoopND}
Expression (\ref{cM1}) for the complexity integral can be notably simplified in case of single loop models.
The single loop $\ell$ of length $\vert \mathcal{M}\vert$ reduces (\ref{eq:chis}) to $\chi_{\mu,\nu}=\chi_{\mu}=\chi$ and $\chi=\prod_{\mu\in \ell} \gamma_{\mu}$. Enforcing these relations into the determinant of the FIM (\ref{eq:detJ2}) yields
\begin{equation}
\begin{split}
\det \mathbb{J}(\vecg) = \prod_{\mu\in\ell} \Bigl(1-\gamma_{\mu}^2\Bigl) \Biggl[\,\prod_{\mu\in\ell} \Bigl( 1 - \frac{\chi}{1 + \chi}\frac{1+\gamma_{\mu}^2}{\gamma_{\mu}^2}\Bigl)\Biggl]\bigl( \det (\matId +\mathbb{B}) \bigl) \\
B_{\mu\nu} = \frac{\gamma_{\mu}}{\gamma_{\nu}} \frac{1-\gamma_{\nu}^2}{1 + \chi} \frac{\chi}{(\gamma_{\mu}^2  - \chi)}
\end{split}
\label{eq:detJ1Loop}
\end{equation}
The matrix $\mathbb{B}(\vecg)$ in (\ref{eq:detJ1Loop}) can be rewritten as the outer product of two vectors:
\begin{equation}
\mathbb{B}(\vecg)=\mathbf{b}\mathbf{c}^t \qquad \text{with}\qquad b_{\mu}=\frac{\chi}{1+\chi}\frac{\gamma_{\mu}}{\gamma_{\mu}^2-\chi}\quad \text{and}\quad\,c_{\mu}=\frac{\alpha_{\mu}(1-\gamma_{\mu}^2)}{\gamma_{\mu}}\,,\, \mu\in\ell
\end{equation}
such that 
\begin{eqnarray}
\det (\matId +\mathbb{B})\,=\, 1+\mathbf{c}^t\mathbf{b}\,=\, 1+\frac{\chi}{1+\chi}\sum_{\mu\in\ell}\frac{1-\gamma_{\mu}^2}{(\gamma^2-\chi)}.
\label{eq:det1Bloop}
\end{eqnarray}
The resulting expression for the complexity, obtained by replacing (\ref{eq:det1Bloop}) and (\ref{eq:detJ1Loop}) in Eq.~(\EqFIMm) of the main text,
\begin{equation}
    \label{cM1LoopND}
  e^{c_\ell}=\pi^{\vert \mathcal{M}\vert}\int_{\left[-1,1\right]^{\vert \mathcal{M}\vert} }\dd\boldsymbol{\gamma}\,q(\boldsymbol{\gamma})\Biggl[\prod_{\mu} \sqrt{\Bigl( 1 - \frac{\chi}{1 + \chi}\frac{1+\gamma_{\mu}^2}{\gamma_{\mu}^2}\Bigl)}\,\Biggl]\sqrt{ 1+\frac{\chi}{1+\chi}\sum_{\mu\in\ell}\frac{1-\gamma_{\mu}^2}{(\gamma_{\mu}^2-\chi)}}\,,
\end{equation}
is then suited for numerical integration using Monte Carlo methods as explained in section {\ref{app:complex_general}}.\\

Fig.~\FigThree~ of the main text displays the complexity of models with a fixed number $|\mathcal{M}|$ of operators and a single loop $\ell$ of length~$k$, 
for different values of $k$.
Such a model has $|\mathcal{M}|-k$ {\it free} operators (operators not involved in any loop).
It can be, for instance, a model with $(|\mathcal{M}|-1)$ fields (considering that the number of spins $n\geq |\mathcal{M}|-1$) and one $(k-1)$-body interaction, or a model formed with a closed chain of $k$ pairwise interactions and $|\mathcal{M}|-k$ free fields.
The complexity of such a model is $c_{\mathcal{M}}(k)=c_{\ell}(k)+(|\mathcal{M}|-k)\log\pi$, where $(|\mathcal{M}|-k)\log\pi$ is the contribution of the free operators and 
$c_{\ell}(k)$ is the one of the single loop of length $k$.
For $k=3$, this latter complexity can be obtained analytically from ~\eqref{cM1LoopND}, 
\begin{align}
    c_{\ell}(3) 
        = \log\int_{\left(-1,1\right)^{3} }
            \frac{1}
                {(1 + x y z)^2}\;\dd x\dd y\dd z 
        = \log(\pi^2)\,,
\end{align}
or directly 
by setting $n=2$ in ~\eqref{cMcompA} (as the complete model for $n=2$ spins is a single loop of length $k=3$).
We thus obtain that $c_{\mathcal{M}}(3)=(|\mathcal{M}|-1)\log\pi$.  
For larger values of $k$, the complexity $c_{\ell}(k)$ 
is obtained numerically by integrating ~\eqref{cM1LoopND} with a Monte Carlo method.
Fig.~\FigThree~ of the main text shows that the complexity $c_{\mathcal{M}}(k)$ of such models increases with the length $k$ of the loop, and saturates for large $k$ at $c_{\mathcal{M}}(k) \rightarrow c_{\mathcal{M}}(3)+\log\pi =|\mathcal{M}|\log\pi$, which corresponds to the complexity of a model with $|\mathcal{M}|$ independent operators.
This can be re-written in term of the complexity of the single loop: the complexity of the single loop increases with the length $k$ of the loop, starting from $c_{\ell}(3)=2\log\pi$ to finally grow for large $k$ as $c_{\ell}(k)\rightarrow k\log\pi$, as if the $k$ operators of the loop were independent.


\subsection{All models for $n\le 4$} \label{app:n=4}
In Table~\ref{Table:n3} and Table~\ref{Table:n4}, we classified all the models for, respectively, $n=3$ and $n=4$.\\

In the $4$-spin system, there are $2^{15}~=~32 768$ distinct non-degenerate models. We counted $20 160$ possible gauge transformations, 
which is in agreement with Eq.~(\ref{numerotrasfgauge}).
By applying all gauge transformations to all models, we find that models can be classified in $46$ complexity classes (see Table~\ref{Table:n4}). 
The number of different values of complexity $c_{\mathcal{M}}$ to be estimated numerically with Eq.~(\ref{cM1LoopND}) is thus drastically reduced, 
from $32 768$ to only $46$.
In the 3-spin system, there are $2^7=128$ non-degenerate models spread over $10$ complexity classes (see Table~\ref{Table:n3}). 
Note that the $3$-spin system is a sub-case of a $4$-spin system (see all classes with $n_{\mathcal{M}}\leq 3$ in Table~\ref{Table:n4}): every complexity class 
of the $3$-spin system is also a class of the $4$-spin system, with all characteristics preserved,  
except for the number of elements in the class (which takes into account the additional spin).\\

The comprehensive study of classes for $n=3$ and $n=4$ allows comparing with the results of the previous sections. 
First, the relation between $|\mathcal{M}|$, $\lambda$ and $n_{\mathcal{M}}$ in ~\eqref{eq:lambda_M} is always verified for each class of the two tables.
We can also observe, in both tables, the ``symmetry'' in the cardinality between classes of models with $|\mathcal{M}|$ operators and their respective complementary classes of models with $2^n-1-|\mathcal{M}|$ operators (see~\ref{app:complementary_models}).
~\eqref{eq:indep_param:N_ind} for the cardinality of classes with only independent operators (such that $|\mathcal{M}|=n_{\mathcal{M}}$) is also verified here.
Finally, let us remark, in Table~\ref{Table:n3}, that two different classes, with different structures (even different number of operators), may have the same 
value of $c_{\mathcal{M}}$. For instance, the model $\mathcal{M}=\{s_1, s_2,s_3\}$ (in the class of the 4th row) and the model 
$\mathcal{M}=\{s_1, s_2, s_3, s_2s_3\}$ (class in the 7th row) have both $c_{\mathcal{M}}=3\log \pi$ even though their structure are clearly different.\\

Remarkably, the complexity of models is not monotonic in the number $|\mathcal{M}|$ of operators, as it can be verified 
in Table~\ref{Table:n3} for $n=3$ and in Fig.~\FigFour~ for $n=4$.
Observe for instance in Table~\ref{Table:n3} that the complete model (with $\vert \mathcal{M} \vert=7$) is much less complex then any model 
with $\vert \mathcal{M} \vert=6$ operators.
At equal number of operators $|\mathcal{M}|$, the maximum of the complexity is achieved by models with only independent operators (see~\ref{SM-4.2} and Fig.~\FigFour~);
on the other hand, \textit{sub-complete models}, i.e. models that contain a \textit{complete} model (see \ref{SI-5}) on a subset of spins,
are the simplest. 
We also notice that complexity decreases when turning an independent interaction into an operator that enter in a loop (compare for instance the two complexity classes with $|\mathcal{M}| = 3$ operators in Table \ref{Table:n3}).
In summary, we found that, adding a new operator to a model will:\\[0.5mm]
    \indent $\circ$ increases its complexity $c_{\mathcal{M}}$ by $\log\pi$ if this new operator is independent (doesn't enter in any loop)\\[0.5mm]
    \indent $\circ$ increases its complexity by a quantity between $0$ and $\log\pi$ if this new operator enter in a single loop;
    $0$ if the length of the loop is $|\ell|=3$, and then growing values for larger loop length (see~\ref{App:SingleLoopND}).\\[0.5mm]
    \indent $\circ$ if the new operator enter in several loops, the complexity may increase (from always less than $\log\pi$) or decrease;
    it is no trivial to predict what will happen, however we observe that, at fix number of operators $|\mathcal{M}|$, the closest the model is to a {\it sub-complete model}, the less complex it is.\\

\begin{table}[h]
\begin{center}
\begin{tabular}{|c|c|c|c|c|c|c|}\hline
$~|\mathcal{M}|~$ & $~n_{\mathcal{M}}~$
& $~\lambda$ & $\{\,|\ell|\,,\;\forall\; \ell\in\mathcal{L} 
\}$ & number of models & $\exp( c_{\mathcal{M}} )$\\ \hline
$0$ & $0$ & $0$ & $\{0\}$ & $1$ & - \\ \hline
$1$ & $1$ & $0$ & $\{0\}$ & $7$ & $\pi \simeq 3.141$\\ \hline
$2$ & $2$ & $0$ & $\{0\}$ & $21$ & $\pi^2 \simeq 9.869$\\ \hline
$3$ & $3$ & $0$ & $\{0\}$ & $28$ & $\pi^3 \simeq 31.006$\\ \hline
$3$ & $2$ & $1$ & $\{0,3\}$ & $7$ & $\pi^2 \simeq 9.869$\\ \hline
$4$ & $3$ & $1$ & $\{0,4\}$ & $7$ & $\pi^{3.56831}\simeq 59.427$\\ \hline
$4$ & $3$ & $1$ & $\{0,3\}$ & $28$ & $\pi^3 \simeq 31.006$ \\ \hline
$5$ & $3$ & $2$ & $\{0,3^2,4\}$ & $21$  & $\pi^{3.18346} \simeq 38.252$\\ \hline
$6$ & $3$ & $3$ & $\{0,3^4,4^3\}$ & $7$  & $\pi^{3.34058}\simeq 45.790$\\ \hline
$\bf 7$ & $\bf 3$ & $\bf 4$ & $\bf \{0,7,4^7,3^7\}$ & $1$ & $\bf \pi^{2.43472}\simeq 16.233$\\ \hline
\end{tabular}
\end{center}
\caption{Summary table of all non-degenerate models of a 3-spin system; models are partitioned in
$10$ classes. Each line gives the characteristics of one class: the common 
structure of the models of the class (number of operators $|\mathcal{M}|$, number of independent operators $n_{\mathcal{M}}$, $\lambda = \log_2 |\mathcal{L}|$, and lengths $|\ell|$ of each loop $\ell\in\mathcal{L}$), 
the number of models in the class, and finally, the corresponding value of the complexity $c_{\mathcal{M}}$.
The notation $a^b$ means that the element $a$ is repeated $b$ times. The last row corresponds to the {\it complete} $3$-spin model.
}
\label{Table:n3}
\end{table} 
\begin{table}[h]
\begin{center}
\begin{tabular}{|c|c|c|c|c|}\hline
~$|\mathcal{M}|$~~ & number of classes & $n_{\mathcal{M}}$ & $\lambda$ & cardinality of each class\\ \hline
$0$ & $1$ & $\{0\}$ & $\{0\}$ & $\{1\}$ \\ \hline
$1$ & $1$ & $\{1\}$ & $\{0\}$ & $\{15\}$ \\ \hline
$2$ & $1$ & $\{2\}$ & $\{0\}$ & $\{105\}$ \\ \hline
$3$ & $2$ & $\{3, 2\}$ & $\{0, 1\}$ & $\{420,35\}$ \\ \hline
$4$ & $3$ & $\{4, 3^2\}$ & $\{0,1^2\}$ & $\{840,420,105\}$ \\ \hline
$5$ & $4$ & $\{4^3, 3\}$ & $\{1^3,2\}$ & $\{1680,840,168,315\}$ \\ \hline
$6$ & $5$ & $\{4^4, 3\}$ & $\{2^4,3\}$ & $\{2520,420,1680,280,105\}$ \\ \hline
$\bf 7$ & $6$ & $\{4^5, \bf 3\}$ & $\{3^5,\bf 4\}$ & $\{840,120,2520,2520,420,\bf 15\}$ \\ \hline
$8$ & $6$ & $\{4^6\}$ & $\{4^6\}$ & $\{840,120,2520,2520,420,15\}$ \\ \hline
$9$ & $5$ & $\{4^5\}$ & $\{5^5\}$ & $\{2520,420,1680,280,105\}$ \\ \hline
$10$ & $4$ & $\{4^4\}$ & $\{6^4\}$ & $\{1680,840,168,315\}$ \\ \hline
$11$ & $3$ & $\{4^3\}$ & $\{7^3\}$ & $\{840,420,105\}$ \\ \hline
$12$ & $2$ & $\{4^2\}$ & $\{8^2\}$ & $\{420,35\}$ \\ \hline
$13$ & $1$ & $\{4\}$ & $\{9\}$ & $\{105\}$ \\ \hline
$14$ & $1$ & $\{4\}$ & $\{10\}$ & $\{15\}$ \\ \hline
$15$ & $1$ & $\{4\}$ & $\{11\}$ & $\{1\}$ \\ \hline
\end{tabular}
\end{center}
\caption{Summary table of 
all non-degenerate models of a 4-spin system; models are partitioned in $46$ classes.
Each line corresponds to one or more classes with the same number of operators $|\mathcal{M}|$ and gives, for each class, the following characteristics:
number of independent operators $n_{\mathcal{M}}$, cardinality of the minimal generating set of loops $\lambda = \log_2{|\mathcal{L}|}$, and number of models in the class (column ``cardinality'').
The notation $a^b$ means that the element $a$ is repeated $b$ times.
The complexity class taken as an example in Fig.~\FigOne~ and Fig.~\FigFour~ Left is highlighted in bold in this table 
($|\mathcal{M}|=7$ interactions, $\lambda = 4$ and a cardinality of $15$).
}
\label{Table:n4}
\end{table}

To conclude, our close analysis of the $n=4$ spin case suggests that the simplest models are those where operators concentrate their support on a subset of spins (and their equivalent models), 
as opposite to ``sparse'' models with many independent parameters. 
More precisely, for fixed value of $|\mathcal{M}|$, classes with lower value of $n_{\mathcal{M}}$ (i.e. with less independent operators) are less complex (see colors in Fig.~\FigFour~).
They are the classes that contains at least one model whose interactions are grouped on a subset $n_{\mathcal{M}}$ of the $n$ spins.
Finally, among these models, the least complex are the ones equivalent to the model that is as close as possible to a {\it sub-complete model}
(exactly a sub-complete model for $|\mathcal{M}|=2^{n_{\mathcal{M}}}-1$, see  $|\mathcal{M}|=1$, $3$, $7$ and $15$ in Fig.~\FigFour~).

\section{Degenerate models}\label{SI-7}
In degenerate models at least two operators, say $\phi^{\mu}$ and $\phi^{\nu}$, have the same parameter, i.e. $g^{\mu}=g^{\nu}$ (see \ref{SI-02}). Since the mapping between parameters and operators is no longer bijective, for a degenerate model $\mathcal{M}$, together with the $|\mathcal{M}|$ operators $\boldsymbol{\phi}$ and $m$ parameters $\vecg$, one requires to specify the matrix $\matQ$ (\ref{eq:matrixQ}), that maps operators into parameters~\footnote{Notice that here we keep indicating the number of operators with $|\mathcal{M}|$, even if for degenerate models it doesn't refer to the cardinality of a set.}.
The degeneracy coefficient $\alpha_{i}=\sum_j U_{ij}$ defines the number of operators parametrized by $g^{i}$, implying that, if the number of parameters is  $m$ , while the number of operators is $\vert \mathcal{M} \vert$, then $\sum_{i=1}^m \alpha_i=\vert \mathcal{M} \vert$. 

The partition function (see \ref{suppZ}) of a generic degenerate spin model $\mathcal{M}$, parametrized by $\vecg$ is
\begin{equation}
 Z_{\mathcal{M}} (\vecg) = 2^n  \prod_{j=1}^m (\cosh g^j)^{\alpha_j}  \sum_{\mathcal{\ell}\in \mathcal{L_{\mathcal{M}}}} \prod_{i \in \mathcal{\ell}} (\tanh g^i)^{\beta_i(\mathcal{\ell})}\,.
 \label{eq:genZD}
\end{equation}
This expression  extends Eq.~(\Eqpartf) of the main text to the case of degenerated parameters. Here  $i \in \ell$ means that there is at least one operator $\phi^{\mu}$, parametrized by $g^i$, such that $\phi^{\mu}$ enters the loop $\ell$ ($\mu \in \ell$ following the notation of \ref{SI-2.1}).  
Finally $\beta_i(\mathcal{\ell})$~\footnote{For a loop $\protect\ell=\{\phi^{\mu_{1}},...,\phi^{\mu_{|\ell|}}\}$ one can construct the matrix $\protect\matQ^{\ell}=\{\vecQ_{:\mu_{1}},...,\vecQ_{:\mu_{|\ell|}}\}$, out of the columns of $\matQ$ in (\ref{eq:matrixQ}), that maps the operators in $\ell$ to their parameters. It follows that $\beta_i(\ell)=\protect\sum_{j=1}^{|\ell|} U^{\ell}_{i\mu_j}$ are the degeneracy coefficients of the parameters in the submodel $\ell$.} denotes the degeneracy coefficient of $g^i$ in loop $\mathcal{\ell}$ (how many operators parametrized by $g^i$ enter the loop  $\mathcal{\ell}$).

\subsection{Independent operators and parameters}\label{SI-7.1}

The partition function of a model with $n_\mathcal{M}=\vert \mathcal{M} \vert$ independent operators and $m$ parameters with degeneracy coefficients $(\alpha_1, \dots, \alpha_m)$ contains only the first term of (\ref{eq:genZD}), $Z_{\mathcal{M}}(\vecg)= 
 2^n \prod_{i=1}^m\cosh(g^i)^{\alpha_i}$. As a consequence its complexity is: 

\begin{equation}
e^{c_{\mathcal{M}}} = \pi^{m}\,\prod_{i=1}^m\sqrt{\alpha_i}.
\end{equation} 
such that any parameter $g^i$ contributes with a 
term $\log (\sqrt{\alpha_i}\pi)$ to the complexity $c_{\mathcal{M}}$. 

Consider now a generic degenerate model $\mathcal{M}$ with $\vert \mathcal{M}\vert$ operators, including $K$ independent $\phi^1, \dots\phi^K$. We can then introduce a model $\mathcal{M}^\prime$ formed by the set of non-independent operators of $\mathcal{M}$ (and relative parameters), and a model $\mathcal{M}_{\rm ind}=\{\phi^1, \dots\phi^K\}$ (and relative parameters). Suppose now that the set of parameters of the two models $\mathcal{M}^\prime$ and $\mathcal{M}_{\rm ind}$ is disjoint, meaning that there is no parameter parametrizing both operators in $\mathcal{M}^\prime$ and $\mathcal{M}_{\rm ind}$. It follows that $\mathcal{M}=\mathcal{M}^\prime\cup\mathcal{M}_{\rm ind}$, such that, analogously to the non-degenerated case (see \ref{SI-4.1}), the partition function (\ref{eq:genZD}) can be factorized in the two models, and so does the complexity:

\begin{align}
    e^{c_{\mathcal{M}}}
        = e^{c_{\mathcal{M}_{\rm ind}}}\;\times\;e^{c_{\mathcal{M}^\prime}} 
        = \pi^{m^{\rm ind}}\prod_{i=1}^{m^{\rm ind}}\sqrt{\alpha_i}\;\times\;e^{c_{\mathcal{M}^\prime}}\,.
\end{align}
where $m^{\rm ind}$ is the number of parameters in $\mathcal{M}_{\rm ind}$ (that possibly differs from the number of operators in $\mathcal{M}_{\rm ind}$, $K$).

 Notice that if the sets of operators of $\mathcal{M}^\prime$ and $\mathcal{M}_{\rm ind}$ are disjoint while the sets of parameters aren't, the complexity doesn't factorize in the two models. By comparing this result with the non degenerate case (see SI-4) degenerating parameters reduces the number of independent operators of a model, decreasing the complexity of the model itself. For models with only independent parameters this statement can be easily checked, as the complexity of a model with $\vert \mathcal{M} \vert$ independent operators and $\vert \mathcal{M} \vert$ parameters is larger than the complexity of a model $\vert \mathcal{M} \vert$ independent operators and $m\leq\vert \mathcal{M} \vert$ parameters, since $\pi^{\vert \mathcal{M} \vert}\leq \pi^{m}\prod_{i=1}^m {\sqrt{\alpha_i}}$ if $\sum_{i=1}^m \alpha_i=\vert \mathcal{M} \vert$. The fact that degeneracy of parameters reduces the complexity of a model holds also in loopy models, as we show in section \ref{App:SingleLoop}.
 
 \subsubsection*{\textbf{Complexity classes}}

In case of degenerated parameters the complexity of models with $n_{\mathcal{M}}=\vert \mathcal{M} \vert$ independent operators depends only on the number of parameters  and the degeneracy coefficients $\alpha_i$, as shown in \ref{SI-7.1}. Specifically each class of complexity is identified by the number of parameters $m\leq \vert \mathcal{M} \vert $, the set of degeneracy coefficients and their multiplicities $\left\lbrace (\alpha_{j_{1}},r_{j_{1}}),...,(\alpha_{j_{K}},r_{j_{K}})\right\rbrace $, with $\alpha_{j_{1}}<...<\alpha_{j_{K}}$ and $K$ total number of distinct values that the degeneracy coefficients $\alpha_i$ take, such that $\sum_{i=1}^K \alpha_{j_{i}}r_{j_{i}}=\vert \mathcal{M} \vert$ and $\sum_{i=1}^K r_{j_{i}}=m$. Then the cardinality of such a class of complexity is

\begin{equation}
\mathcal{N}_{\rm ind}^{deg} = \mathcal{N}_{\rm ind}\frac{n_{\mathcal{M}}!}{(\alpha_{j_1}!)^{r_{j_1}}\cdot ... \cdot(\alpha_{j_K}!)^{r_{j_K}}}\frac{1}{r_{j_1}!\cdot ... \cdot r_{j_K}!}
\end{equation}
the number of possible ways of choosing $n_{\mathcal{M}}$ independent operators $\mathcal{N}_{\rm ind}$ (\ref{eq:indep_param:N_ind}) times the number of partitions of a set of $n_{\mathcal{M}}$ elements in exactly $m$ subsets of which $r_{j_1}$ of cardinality $\alpha_{j_1}$, ..., and 
$r_{j_K}$ of cardinality $\alpha_{j_K}$.
 
 \subsection{Generic model}\label{app:complex_genericDM}

The derivation of the complexity for the generic model in case of non-degenerated parameters (see \ref{app:complex_genericND}) can be straightforwardly extended to the case of degenerate models. In particular the FIM (hessian of logarithm of the partion function (\ref{eq:genZD})) is

\begin{equation}
J_{ij} = \begin{cases}
  \Bigl(1 - \gamma_i^2\Bigl)\Bigl(\alpha_i+ \frac{(1-\gamma_i^2)}{\gamma_i^2(\chi+1)}\left[\chi_{i,i}-\chi_i \right]  - 2 \frac{\chi_i}{1 + \chi} - \frac{1 - \gamma_i^2}{\gamma_i^2} \frac{\chi_i^2}{(1+\chi)^2}\Bigl) &\text{for}\qquad i=j\\ 
 &\\
  \frac{1 - \gamma_i^2}{\gamma_i (1 + \chi)} \frac{1 - \gamma_j^2}{\gamma_j (1 + \chi)} \Bigl(\chi_{i,j} (1 + \chi) - \chi_i \chi_j\Bigl)&\text{for}\qquad i\ne j
\end{cases}
\label{eq:FisherDM}
\end{equation}
where
\begin{subequations}
\begin{eqnarray}
\gamma_i &=& \tanh g^i  \\
\chi &=& \sum_{\mathcal{\ell}\in \mathcal{L_{\mathcal{M}}}\setminus\{\emptyset\}} \prod_{i \in \mathcal{\ell}} \gamma_i^{\beta_i(\mathcal{\ell})}\\
\chi_i &=&  \sum_{\mathcal{\ell} | i } \beta_i(\mathcal{\ell})\prod_{j \in \mathcal{\ell}} \gamma_j^{\beta_j(\mathcal{\ell})} \\
\chi_{i,j} &=&  \sum_{\mathcal{\ell} | \{i,j\} }\beta_i(\mathcal{\ell})\beta_j(\mathcal{\ell}) \prod_{m \in \mathcal{\ell}} \gamma_m^{\beta_m(\mathcal{\ell})}
\end{eqnarray}
\label{eq:chisD}
\end{subequations}
and $\mathcal{\ell} | i $ and \ $\mathcal{\ell} | \{i,j\} $ refer respectively to the loops in $\mathcal{L_{\mathcal{M}}}$ in which $g^i$ enters and in which both $g^i$ and $g^j$ enter.
The derivation then follows the non degenerated case, by decomposing the FIM (\ref{eq:FisherDM}) and factorizing out a diagonal matrix ( see \ref{app:complex_genericND}).

Finally the complexity of the generic degenerate model reads:

\begin{equation}
    \label{cM1DM}
  e^{c_\mathcal{M}}=\pi^{m}\int_{\left[-1,1\right]^m }\dd\boldsymbol{\gamma}\,q(\boldsymbol{\gamma})\Biggl[\prod_{i} \sqrt{\Bigl( \alpha_i - \frac{\chi_i}{1 + \chi}\frac{1+\gamma_i^2}{\gamma_i^2}\Bigl)}\,\Biggl]\sqrt{ \det (\matId +\mathbb{B})}
\end{equation}
where the matrix $\mathbb{B}$ is defined as
\begin{equation}
B_{ij} = \frac{\gamma_i}{\gamma_j} \frac{1-\gamma_j^2}{1 + \chi} \frac{\chi_{i,j}(1 + \chi) - \chi_i \chi_j}{(\alpha_i\gamma_i^2 (1 + \chi) - (1+\gamma_i^2)\chi_i)}
\end{equation}
and
\begin{equation}
q(\boldsymbol{\gamma})=\frac{1}{\pi^{m}}\prod_{i} \Bigl(1-\gamma_i^2\Bigl)^{-1/2}\,.
\label{eq:measureq} 
\end{equation}

Now (\ref{cM1DM}) is prone for standard Monte Carlo integration by random sampling $\boldsymbol{\gamma}$ according to the pdf $q(\boldsymbol{\gamma})$. 

\subsection{Models with a single loop} 
\label{App:SingleLoop} 
We now focus on the complexity of models that contain only one loop involving all $\vert \mathcal{M}\vert$ operators, $\mathcal{L}=\{\{\emptyset\}, \ell\}$, where $\ell=\{\phi^1, \dots, \phi^{\vert \mathcal{M}\vert}\}$, parametrized by $m\leq \vert\mathcal{M}\vert$ possibly degenerated parameters. All such models with fixed number of operators, parameters and degeneracy coefficient belong to the same class of complexity. Specifically the class of complexity is identified by the number of operators $\vert\mathcal{M}\vert$, the number of parameters $m$ and the set of degeneracy coefficients and their multiplicities $\left\lbrace (\alpha_{j_{1}},r_{j_{1}}),...,(\alpha_{j_{K}},r_{j_{K}})\right\rbrace $, with $\alpha_{j_{1}}<...<\alpha_{j_{K}}$ ( $K$ is the total number of distinct values that the degeneracy coefficients $\alpha_i$ take), such that $\sum_{i=1}^K \alpha_{j_{i}}r_{j_{i}}=\vert \mathcal{M} \vert$ and $\sum_{i=1}^K r_{j_{i}}=m$. The cardinality of such a class of complexity is

\begin{equation}\label{eq:NumSingleLoopDeg}
\mathcal{N}_{\rm 1\, loop}^{deg} = \mathcal{N}_{\rm 1 \, loop}\frac{\vert \mathcal{M} \vert!}{(\alpha_{j_1}!)^{r_{j_1}}\cdot ... \cdot(\alpha_{j_K}!)^{r_{j_K}}}\frac{1}{r_{j_1}!\cdot ... \cdot r_{j_K}!}
\end{equation}
given by the number of possible ways of choosing $\vert \mathcal{M} \vert$ operators constituting a single loop $\mathcal{N}_{\rm 1 \, loop}$ (\ref{eq:NumSingleLoop}), times the number of partitions of a set of $\vert \mathcal{M} \vert$ elements in exactly $m$ subsets of which $r_{j_1}$ of cardinality $\alpha_{j_1}$, ..., and 
$r_{j_K}$ of cardinality $\alpha_{j_K}$.\\

The complexity of this class of models can be derived from (\ref{cM1DM})
by enforcing the single loop constraints on (\ref{eq:chisD}), namely $\alpha_i=\beta_i$ and $\chi_i=\alpha_i\chi$, $\chi_{i,j}=\alpha_i\alpha_j\chi$, while $\chi=\prod_{i=1}^m \gamma_i^{\alpha_i}$: 
\begin{equation}
    \label{cM1DMSingleLoop}
  e^{c_\mathcal{M}}=\pi^{m}\int_{\left[-1,1\right]^m }\dd\boldsymbol{\gamma}\,q(\boldsymbol{\gamma})\Biggl[\prod_{i} \sqrt{\Bigl( \alpha_i - \frac{\chi}{1 + \chi}\frac{1+\gamma_i^2}{\gamma_i^2}\Bigl)}\Biggl]\sqrt{ 1+\frac{\chi}{1+\chi}\sum_i\alpha_i\frac{1-\gamma_i^2}{(\gamma^2-\chi)}}
\end{equation}
where $q(\boldsymbol{\gamma})$ is defined in ~\eqref{eq:measureq}. 
The expression (\ref{cM1DMSingleLoop}) for the complexity was obtained by replacing the relation (that only holds for single loop models)
\begin{equation}
\det (\matId +\mathbb{B})= 1+\frac{\chi}{1+\chi}\sum_i\alpha_i\frac{1-\gamma_i^2}{(\gamma^2-\chi)}.
\end{equation}
in the complexity of the generic degenerate model (\ref{cM1DM}).\\

\begin{figure}[ht]
\centering
\includegraphics[width=0.5\columnwidth]{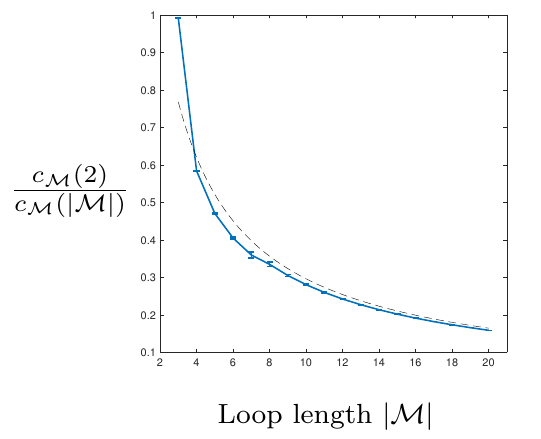}
\caption{\label{fig:Degeneracy} Degeneracy and complexity.
Ratio (data points + solid line to guide the eye) between the complexity of two single loop models of length $|\mathcal{M}|$ (number of operators) parametrised respectively by $2$ parameters (one of them $|\mathcal{M}|-1$ times degenerated) and $|\mathcal{M}|$ parameters, versus loop length $|\mathcal{M}|$. Complexities here are averages of $10^3$ numerical estimates of the integrals (\ref{cM1LoopND}) and (\ref{cM1DMSingleLoop}) using $10^6$ MC samples each and error bars result from error propagation of one standard deviation of these estimates. The larger the loop the more independent are the operators, as shown in Fig.~\FigThree~ of the main text, such that the ratio between the complexities of single loop degenerated and non degenerated models is approaching the ratio (dashed line) between the complexities of independent operators degenerated ($\log \sqrt{|\mathcal{M}|-1}\pi^2$) and non-degenerated ($\log\pi^{|\mathcal{M}|}$) models (see \ref{SI-4.1} and \ref{SI-7.1}).
}
\end{figure}

The simple case of single loop models constitutes a suitable platform to gain some insights on how the degeneracy affects the complexity of models with loops. The degeneracy coefficients of parameters enter the expression of the complexity (\ref{cM1DMSingleLoop}) in a non trivial way, such that numerical exploration is required. In Fig.~\ref{fig:Degeneracy} we compare the complexity of a single loop model of length $|\mathcal{M}|$ (number of operators) parametrised by  $2$ parameters and the corresponding non degenerated single loop model. By increasing the length of the loop the degeneracy increases while the complexity decreases (relatively to the non degenerated model). The fact that --- analogously to the independent operators model (see \ref{SI-7.1}) --- degeneracy in the single loop model reduces the complexity can be intuitively understood through our general argument in \ref{SI-4}. By degenerating the parameters $g^\mu$ one indirectly constrains the averages of the operators $\phi^{\mu}$ (dual coordinates $\ensavg^\mu$ in the model manifold) resulting in a downsized marginal polytope $\mathcal{F}$ and a smaller complexity as a consequence.  

\newpage
\bibliography{pnas-sample}